\documentclass[11pt]{article}
\usepackage[final]{acl}
\usepackage{times}
\usepackage{latexsym}

\usepackage{microtype}
\usepackage[T1]{fontenc}
\usepackage[utf8]{inputenc}
\usepackage[nohints]{minitoc}
\usepackage{graphicx}
\usepackage{xcolor}
\usepackage{amsmath}
\usepackage{amsfonts}
\usepackage{amssymb}
\usepackage{amsthm}
\usepackage{mathrsfs}
\usepackage{mathtools}
\usepackage{bm}
\usepackage{dsfont}
\usepackage{upgreek}
\usepackage{caption}
\usepackage{subcaption}
\usepackage{wrapfig}
\usepackage{booktabs}
\usepackage{multirow}
\usepackage{diagbox}
\usepackage{makecell}
\usepackage{bigstrut}
\usepackage{arydshln}
\usepackage{adjustbox}
\usepackage{algorithm}
\usepackage{algpseudocode}
\usepackage{tikz}
\usepackage{pgf-pie}
\usepackage{minted}
\usetikzlibrary{arrows.meta, positioning, shapes, fit, backgrounds, calc}
\usepackage{cleveref}
\usepackage{url}

\title{SpecAgent: A Speculative Retrieval and Forecasting Agent\\for Code Completion}

\author{
  George Ma\textsuperscript{1}\thanks{Work done as an intern at Amazon Web Services.},
  Anurag Koul\textsuperscript{2},
  Qi Chen\textsuperscript{2},
  Yawen Wu\textsuperscript{2},
  Sachit Kuhar\textsuperscript{2},
  Yu Yu\textsuperscript{2}\\[4pt]
  \textbf{Aritra Sengupta\textsuperscript{2}}\thanks{Correspondence to: \texttt{george\_ma@berkeley.edu}, \texttt{aritras@amazon.com}.}, 
  \textbf{Varun Kumar\textsuperscript{2}}, 
  \textbf{Murali Krishna Ramanathan\textsuperscript{2}}\\[6pt]
  \textsuperscript{1}UC Berkeley \quad
  \textsuperscript{2}Amazon Web Services \\
}

\begin{document}

\maketitle
\doparttoc
\faketableofcontents

\begin{abstract}
Large Language Models (LLMs) excel at code-related tasks but often struggle in realistic software repositories, where project-specific APIs and cross-file dependencies are crucial. Retrieval-augmented methods mitigate this by injecting repository context at inference time. The low inference-time latency budget affects either retrieval quality or the added latency adversely impacts user experience. We address this limitation with \emph{SpecAgent}, an agent that improves both latency and code-generation quality by proactively exploring repository files during indexing and constructing \emph{speculative context} that anticipates future edits in each file. This indexing-time asynchrony allows thorough context computation, masking latency, and the speculative nature of the context improves code-generation quality. Additionally, we identify the problem of \emph{future context leakage} in existing benchmarks, which can inflate reported performance. To address this, we construct a synthetic, leakage-free benchmark that enables a more realistic evaluation of our agent against baselines. Experiments show that SpecAgent consistently achieves absolute gains of 9--11\% (48--58\% relative) compared to the best-performing baselines, while significantly reducing inference latency.
\end{abstract}

\section{Introduction}

Large Language Models (LLMs) have demonstrated remarkable capabilities in a wide range of software engineering tasks, including code completion \citep{code4me,qwen3}, code editing \citep{grace}, issue resolution \citep{swe-bench}, and automated test generation \citep{titanfuzz}. These successes are largely the result of advances in large-scale model pre-training and fine-tuning techniques \citep{codet5,codebert}, fueled by massive, publicly available code-corpus datasets \citep{thestack}. This has led to significant improvement over benchmarks \citep{codex,program-synthesis,crosscodeeval} that test general programming knowledge , logical reasoning, and problem-solving skills.

However, the majority of these benchmarks evaluate models in isolated settings that do not reflect the complexity of real-world software development \citep{swe-bench}. In practice, software engineers often work within large, evolving, and context-rich codebases where understanding the local context, e.g., private dependencies,  is essential to completing tasks correctly. Without access to such context, such as the definitions of project-specific APIs, cross-file dependencies, or coding conventions, LLMs can produce completions or edits that are syntactically correct but semantically inconsistent with the target repository. As a result, bridging the gap between the general capabilities of LLMs and the repository-specific requirements of real-world engineering tasks remains a critical challenge \citep{swe-bench-illusion}.

One promising approach to this challenge has been to enhance LLMs with information retri\-eval (IR) mechanisms that incorporate repository-specific knowledge during inference. This includes techniques like retrieving similar code snippets using BM25 \citep{bm25}, mining semantically related code via dense embeddings \citep{codesage,repocoder}, and extracting structural metadata from the repository \citep{repograph}. Integrating such retrieved information into the model's prompt has been shown to significantly improve performance \citep{repoformer} on tasks that require awareness of cross-file dependencies, custom function signatures, or domain-specific patterns.
While effective in improving correctness, these retrieval-based strategies require querying large indexes or scanning complex repository structures during online inference, which can introduce substantial latency.

This work is in a low-latency domain of in-IDE code auto-completions---inline completions, typically offered by products such as Cursor \citep{cursor}, Copilot \citep{github_copilot}, Amazon Q \citep{amazon_q_developer}, etc. Our aim is to enhance inline completions user experience by improving model performance through richer context, as well as eliminating inference-time retrieval latency. To achieve this, we develop \emph{SpecAgent}, an agent that constructs a structured context for each file at repository indexing time rather than at inference time. By front-loading these costly operations, we streamline the online phase for faster responses. We leverage a novel insight where the agent \emph{speculates} on future functionalities or issues within files and retrieves additional context that supports building or fixing these anticipated changes in the upcoming developer session.

Our key contributions include the design of a speculative context construction framework that anticipates developer needs by pre-computing both current and likely future code-related information. This context is retrieved asynchronously at indexing, allows thorough context computation saving on inference time latency. Additionally, we also identify a critical limitation in existing code-completion benchmarks, which suffer from \textit{``future-context leakage''} where target function invocations across files inadvertently reveal information about future code, such as function signature, intended usage, etc. To address this issue, we introduce a new benchmark, explicitly crafted to eliminate such leakage and provide a more rigorous evaluation environment. Finally, we develop an oracle agent to establish upper-bound performance metrics on this benchmark and demonstrate that leveraging SpecAgent's context improves the Qwen3-8B \citep{qwen3} model's performance by approximately \textbf{10--11\%} (58\% relative), and the Qwen3-30B-A3B \citep{qwen3} model by \textbf{9--10\%} (48\% relative) compared to strong baselines, with no additional inference-time retrieval latency.

\section{Related Works}

\paragraph{Code Language Models.} Code language models underpin a wide range of software engineering tasks, including code completion \citep{code4me}, debugging \citep{self-debugging}, translation \citep{codescribe}, and issue resolution \citep{swe-bench}. Progress in code completion has spanned data filtering, data synthesis, and both pre-training and post-training strategies. Notable developments include StarCoder-v2 \citep{starcoder2}, which scales diversified high-quality code data, and OpenCoder \citep{opencoder}, which leverages code-related web data with synthetic fine-tuning. Seed-Coder \citep{seedcoder} introduces Long Chain-of-Thought (LongCoT) reinforcement learning to improve reasoning, while Qwen3-Coder \citep{qwen3coder} extends model context to 256K tokens.

\paragraph{Repository-Level Code Completion.} Repository level completion is more challenging than function- or file-level tasks, as it requires holistic reasoning over large codebases. Benchmarks such as RepoBench \citep{repobench} and CrossCodeEval \citep{crosscodeeval} evaluate model performance in this setting. Recent methods enhance repository context during inference: RepoFuse \citep{repofuse} integrates dependency and similarity signals; R2C2-Coder \citep{r2c2-coder} fine-tunes models with diverse context types; and \citet{hierarchical-context-pruning} study context integration and pruning strategies. 

\paragraph{Efficient Context Retrieval for Inline Completion.} Code language models face efficiency constraints from limited context windows and strict latency budgets that preclude full-repository input. Retrieval-augmented generation (RAG) mitigates this by dynamically fetching relevant snippets \citep{rag}, though it introduces significant retrieval overhead \citep{coderag-bench}. To reduce this cost, CoSHC \citep{coshc} applies deep hashing for compact similarity search, while SECRET \citep{secret} uses segmented hashing to accelerate lookups. Complementary work \citep{spencer} distills smaller retrievers to reduce embedding latency. Our approach is orthogonal to these optimizations: we shift retrieval computation from inference to indexing time, reframing the cost–latency trade-off in retrieval-augmented code completion.

\section{Preliminaries}\label{sec: preliminaries}

\subsection{Inline Function Completion}

We study the problem of \emph{inline code completion}  \citep{code4me}, specifically \emph{function completion}, in which the goal is to predict the body of a target function within a partially observed source file. At inference time, a code completion model is provided with: \emph{1) Left context:} the content appearing before the target function in the source file. \emph{2) Right context:} the content appearing after the target function in the source file. \emph{3) Prompt:} the signature and docstring of the target function. \emph{4) Cross-file contexts:} additional code snippets retrieved from other files in the repository. The code completion model generates a prediction for the target function body, which is then evaluated against unit tests (when available) \citep{bigcodebench} or through static metrics such as exact match or edit similarity \citep{benchmarks-and-metrics} .

\subsection{Baseline Retrieval Methods for Cross-File Contexts}

Cross-file contexts can be obtained using various retrieval methods. We summarize three representative approaches below. These are state-of-the-art baselines in inline low latency code auto-completions to the best of our knowledge.

\paragraph{Sparse retrieval.} A classical approach that ranks candidate code chunks from other files by token-level similarity to the query, often using methods such as BM25 \citep{bm25}. The query is typically constructed from the target function's signature, docstring, and potentially parts of its surrounding context.

\paragraph{Dense retrieval.} A learned approach that encodes both the query and candidate code chunks into vector representations using a pre-trained embedding model such as UniXcoder \citep{unixcoder} or CodeSage \citep{codesage}. The candidates are ranked by a similarity measure (e.g., cosine similarity) between their embeddings.

\paragraph{RepoMap.} A structural retrieval method that leverages the API structure of the repository. RepoMap \citep{repomap} extracts and indexes information from all files, such as class names, member functions, function signatures, and variable names. For a given target file, RepoMap retrieves related entities from imported files based on the repository's dependency graph.

\subsection{Indexing-Time vs.\ Inference-Time Retrieval}\label{sec:indexing_vs_inference}

Retrieval methods for code completion differ fundamentally in \emph{when} the relevant computations can be performed.  

\begin{itemize}
    \item \emph{Inference-time retrieval}: Sparse or dense retrieval methods rely on queries constructed from information available when the developer begins authoring a target function, such as its signature, docstring, a program statement inside the function, or surrounding file context. Since this information is unavailable in advance, similarity scores and rankings must be computed at inference time, immediately before invoking code-completion model. This introduces additional latency, which can significantly degrade the responsiveness of interactive code completion systems (up to 9 secs in our experiments). Moreover, as we discuss in \Cref{sec: benchmark_leakage}, existing benchmarks often allow such methods to exploit future context leakage, leading to inflated and unrealistic evaluations.

    \item \emph{Indexing-time retrieval}: In contrast, indexing-time retrieval is performed proactively during repository analysis, before the developer has written the target function. For example, RepoMap~\cite{repomap} builds a repository-wide API map that is fully computable ahead of time, without access to the inference-time query. In this paradigm, agents operate on an \emph{indexing-time repository state}, a snapshot that predates the target function and often its associated callers and tests. Because indexing-time analysis is decoupled from user interaction, agents can conduct extensive exploration, static analysis, and even speculative reasoning without affecting user-perceived latency. The contexts associated with a target file are then determined by these pre-computed structures and relations.
\end{itemize}

\begin{figure*}[htbp]
    \centering
    \includegraphics[width=\textwidth]{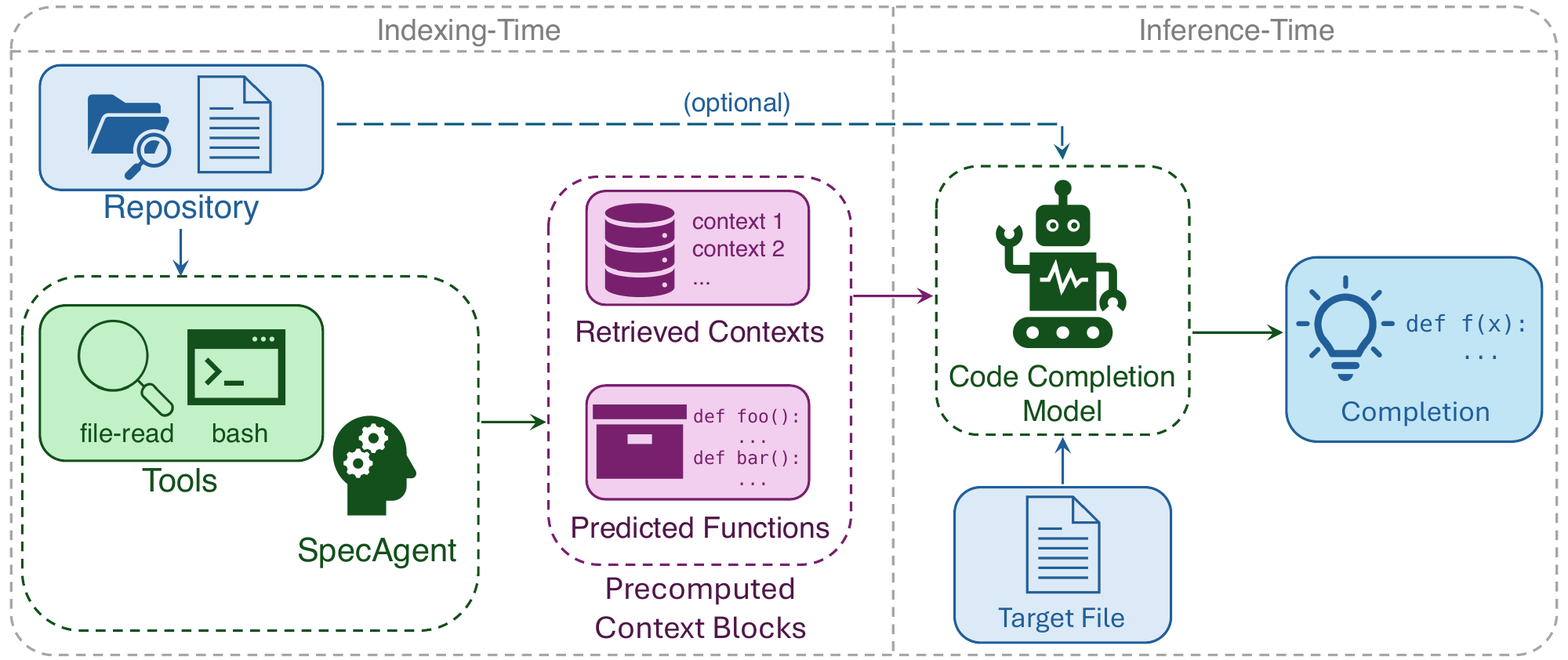}
    \caption{SpecAgent's workflow at indexing-time (left) involves retrieving relevant contexts and generating target function predictions asynchronously. These are then provided to the code completion model during inference (right).}
    \label{fig: indexing_time_state}
\end{figure*}

The distinction between indexing-time retrieval and inference-time retrieval is central to our work. Indexing-time retrieval enables the system to shift costly exploration and context construction away from the latency-critical inference path. This not only improves efficiency, but also allows for richer and more global forms of analysis that would be impractical to execute at inference time. In \Cref{sec: exp}, we show that such indexing-time speculative context construction leads to substantial improvements in model accuracy while introducing no inference-time overhead.

\section{Methodology}\label{sec: method}

This section describes our agent-based approach for constructing cross-file context to support inline code completion. The primary idea is to perform \emph{indexing-time} (background) asynchronous exploration of a repository to produce structured context blocks that are later consumed by a code completion model at \emph{inference time}. We formalize the indexing-time settings and describe the indexing-time agent family (including their interfaces, outputs, and operational constraints). We also propose an oracle agent that serves as a principled upper bound for context quality.

\subsection{Indexing-Time Agents}\label{sec: method_indexing}

We propose a family of agents that, at indexing time, proactively retrieve cross-file contexts and generate target function predictions to enhance code completion.

\paragraph{Indexing-time agent capabilities.} The agent receives the path and content of the target file, along with full-repository access. It can read entire files or specific line ranges, perform keyword searches, and execute read-only shell commands. Prompted to analyze the target file and explore the repository, the agent returns a set of \emph{context blocks}, each representing a specific type of potentially useful information, e.g., related code snippets, dependency structures, interface signatures, error-handling patterns, or speculative implementations. All retrieval is conducted solely on the indexing-time repository state, ensuring no leakage of the ground-truth target function.  

In our experiments (\Cref{sec: exp_main}), each agent produces 12 context blocks per target file. These blocks are later merged with local signals (left file context, right file context, and the prompt) before being passed to the completion model, introducing no additional inference-time latency compared to standard retrieval-based methods.

\paragraph{Agent variants.} We evaluate three complementary variants, each adhering to a common interface and output format:

\begin{itemize}
    \item \emph{Retriever Agent}. It focuses exclusively on \emph{retrieval}. Given the target file, it identifies likely dependencies and usage patterns, searches for relevant code (helpers, call patterns, test snippets), and returns ranked snippets and structural hints to guide the model at inference time. It generates 12 retrieval-based context blocks, due to context-size limitation of inline code-completion model.
    
    \item \emph{Forecaster Agent}. It focuses exclusively on \emph{prediction}. Without retrieving external snippets, it hypothesizes plausible functions a developer might add to the file, generating one or more candidate implementations with brief rationales. It then outputs 12 prediction-based context blocks. The Retriever Agent and Forecaster Agent differ only in their prompting, ensuring comparability across retrieval- and prediction-focused strategies.
    \item \emph{Speculative Agent (SpecAgent)}. It combines retrieval and prediction into a single workflow. SpecAgent constructs a hybrid set of 12 context blocks by taking the top-ranked retrieval blocks from the Retriever Agent and the top-ranked prediction blocks from the Forecaster Agent. This allows the composition of contexts to vary flexibly. In the main experiments (\Cref{sec: exp_main}), we use 9 retrieval blocks and 3 prediction blocks, and we further study alternative compositions in ablation experiments (\Cref{sec: exp_ablation}). By jointly selecting relevant contexts and synthesizing candidate functions, SpecAgent can directly supply accurate completions or high-quality drafts, while retaining the fallback benefits of retrieved evidence.
\end{itemize}

We summarize the indexing-time pipeline of SpecAgent in \Cref{alg: specagent}. For clarity, let $\mathcal{C}(f)$ denote the ordered list of context blocks associated with a target file $f$, and let $k$ denote the total number of blocks (fixed to 12 in our experiments).

\begin{algorithm}[htbp]
    \caption{SpecAgent Indexing-Time Pipeline}
    \label{alg: specagent}
    \begin{algorithmic}[1]
    \Require Repository $\mathcal{R}$, target file $f$, block budget $k$, prediction budget $k_p$, retrieval budget $k_r$
    \Ensure Ordered list of context blocks $\mathcal{C}(f)$ for file $f$
    
    \State Initialize prompts $P_{\text{ret}}$ (retrieval) and $P_{\text{pred}}$ (prediction)
    
    \Statex
    
    \State \textbf{Retriever Agent:}
    \State Invoke agent with prompt $P_{\text{ret}}$ and read-only tools over $\mathcal{R}$
    \State Agent explores repository using arbitrary sequences of tool calls (e.g., search, file inspection)
    \State $\mathcal{R}_f \gets$ ordered list of up to $k_r$ retrieval context blocks
    
    \Statex
    
    \State \textbf{Forecaster Agent:}
    \State Invoke agent with prompt $P_{\text{pred}}$ and read-only tools over $\mathcal{R}$
    \State Agent analyzes repository structure and synthesizes plausible future functions for $f$
    \State $\mathcal{P}_f \gets$ ordered list of up to $k_p$ prediction context blocks
    
    \Statex
    
    \State \textbf{Composition:}
    \State $\mathcal{C}(f) \gets$ concatenate $\mathcal{P}_f$ and $\mathcal{R}_f$ (preserving order)
    \If{$|\mathcal{C}(f)| > k$}
        \State truncate $\mathcal{C}(f)$ to first $k$ blocks
    \EndIf
    
    \Statex
    
    \State Store $\mathcal{C}(f)$ in index for file $f$
    
    \State \Return $\mathcal{C}(f)$
    \end{algorithmic}
\end{algorithm}

\paragraph{Integration with inference.} Context blocks from indexing-time agents are stored and indexed per file. At inference time, the completion model receives: (i) left file context, (ii) right file context, (iii) the prompt, and (iv) the stored cross-file context blocks. Since all exploratory work is completed asynchronously before inference, this design supports richer, more diverse contexts without increasing inference latency. An example of the code completion model's prompt during inference is shown in \Cref{app: prompt_example}.

\paragraph{Rationale and benefits.} Indexing-time exploration enables deep, repository-wide analysis under realistic operational constraints, amortizing computation across many future completions. The Retriever Agent supplies corroborating evidence that disambiguates interfaces; the Forecaster Agent provides fully formed drafts that may be directly adopted as target code; and SpecAgent combines both to maximize the probability of correct completion under limited attention budgets.

\paragraph{Re-indexing and deployment considerations.} In a realistic IDE deployment, indexing-time context construction runs asynchronously and is not triggered by every keystroke or routine save. Instead, cached contexts are refreshed only when repository changes are large enough to materially alter the file's anticipated future functionality or the relevant cross-file contexts, such as substantial edits, addition or removal of major functionalities, or non-local dependency changes. This makes the preprocessing cost amortizable across many future completion requests. Although the one-time indexing cost per file is non-trivial, it is incurred infrequently and entirely off the latency-critical inference path, while the resulting contexts can be reused many times. In addition, indexing-time agents can be instantiated with lower-cost models while still preserving the main benefits of speculative context construction. Overall, this yields a favorable trade-off for inline code completion: moderate asynchronous preprocessing can improve completion quality without adding inference-time latency. Incremental and modular re-indexing remains an important direction for future work.

\subsection{Oracle Agent}\label{sec: method_oracle}

In code completion, many factors influence performance. Since our work focuses on cross-file context, we design an ``oracle'' retrieval agent to estimate the upper bound achievable by improving context quality alone. This agent  operates in the full \emph{inference-time state}, where both the target function and its callers are present. Unlike the \emph{Retrieval Agent}, it also has access to the ground-truth implementation, enabling analysis of calls, dependencies, and error handling to select highly relevant cross-file contexts—though the model never sees the ground-truth directly. To ensure fairness, any block that copies or paraphrases the target function is filtered out. The oracle uses the same repository tools and formatting as the Retrieval agent, enabling direct comparison across strategies.

\paragraph{Why inference-time?} The oracle upper bounds both indexing-time retrieval (like our agent variants) and inference-time methods (e.g., BM25, dense retrieval), which may surface usage patterns or tests of the target function. It thus simulates the best-case inference-time scenario for evaluating how close practical methods come to this ideal.

\begin{figure*}
    \centering
    \includegraphics[width=\textwidth]{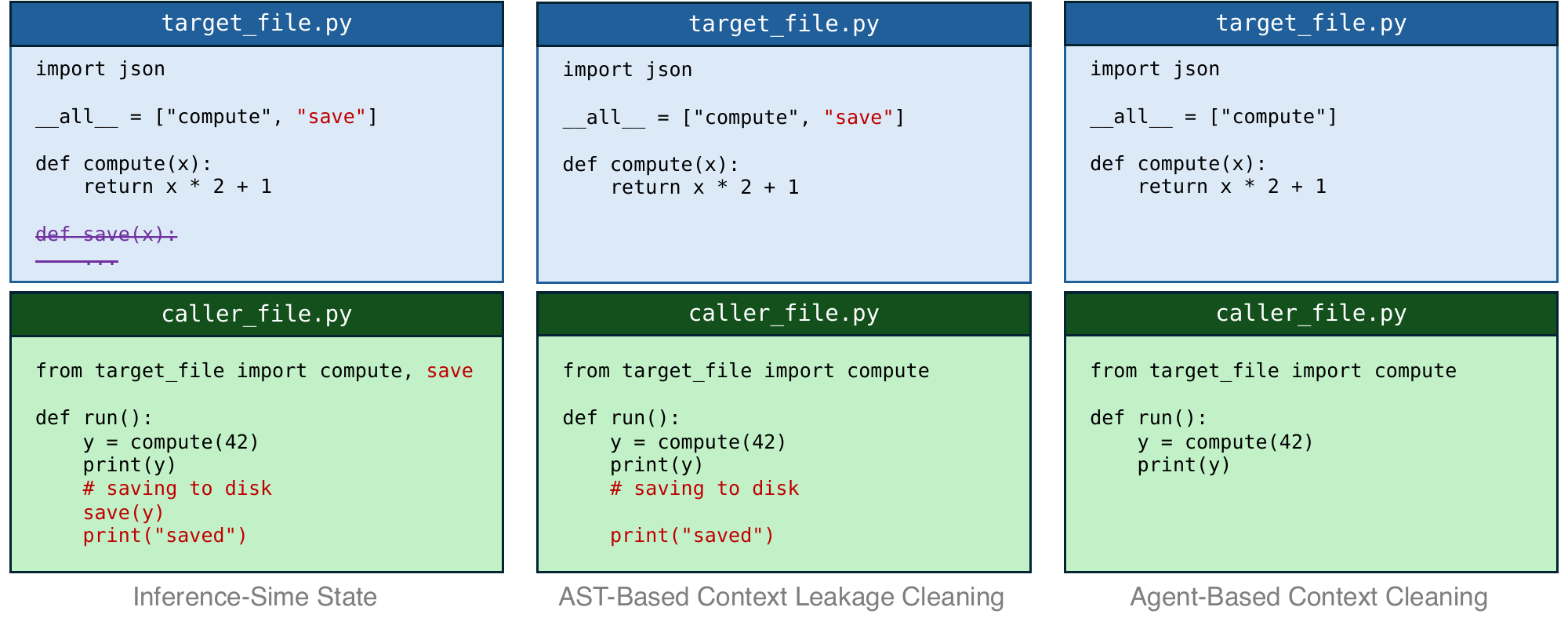}
    \caption{Illustration of future context leakage in existing benchmarks. Although the target function \texttt{save} is removed from \texttt{target\_file.py}, relevant information remains in the target file and \texttt{caller\_file.py}, allowing context retrieval methods to access information that would not exist in a real development scenario.}
    \label{fig: leakage_example}
\end{figure*}

\section{Benchmark}\label{sec: benchmark}

This section presents the benchmark setup used in our study. We begin by identifying a critical flaw in existing code completion benchmarks: \emph{future context leakage}, which makes them incompatible with our experimental goals. We then describe the methodology for constructing a new benchmark specifically tailored for evaluating indexing-time context retrieval.

\subsection{The Future Context Leakage Problem}\label{sec: benchmark_leakage}

Many existing code completion benchmarks (e.g., \citet{repocod, crosscodeeval, repobench}) suffer from a critical flaw in their construction. Given a target function to be completed, these benchmarks typically remove its definition prior to context retrieval. However, this does not eliminate all forms of information leakage: other parts of the repository, such as test files or caller functions, may still reference or depend on the target function. In practice, we observe that retrieval methods can access these leaked code chunks, such as test cases, resulting in artificially inflated performance. This issue contradicts real-world development scenarios, where code that calls a target function would not exist prior to the function's implementation. We refer to this as the \emph{future context leakage} problem, following the terminology introduced by \citet{humanevo}.

We illustrate an example of this problem in \Cref{fig: leakage_example}. As shown in \Cref{fig: leakage_example}, the target function \texttt{save} is defined in \texttt{target\_file.py} and invoked in \texttt{caller\_file.py}. In typical benchmarks, only the definition of \texttt{save} is removed, while \texttt{caller\_file.py} remains untouched. As a result, information about the target function can still be retrieved via its call sites, violating the assumption that the function does not yet exist during context retrieval.

Such leakage leads to misleading evaluations. For existing retrieval methods, including sparse retrieval based on lexical similarity and dense retrieval based on semantic similarity, this can result in retrieval of the very test cases used for evaluation, thereby inflating performance. In our indexing-time context retrieval setting, the problem becomes even more pronounced. Our retrieval agent is capable of analyzing these leaked call sites and accurately inferring the implementation of the target function, yielding unreasonably high performance that does not reflect realistic conditions.

To our knowledge, the only publicly available benchmark that explicitly addresses this issue is HumanEvo \citep{humanevo}. Unfortunately, we were unable to setup their containerization environment. As a result, we construct a synthetic benchmark specifically designed for indexing-time context retrieval without future context leakage. While building such a benchmark from real-world repositories is beyond the scope of this paper, we advocate for the development of more realistic code completion benchmarks in the research community.

\subsection{Benchmark Construction}\label{sec: benchmark_creation}

Given a target function and its associated inference-time repository, our goal is to synthesize a plausible state of the repository from an earlier point in time, before the target function has been implemented. This reflects a realistic scenario in which the user has not yet begun authoring the target function, and the retrieval agent operates asynchronously during indexing time. As illustrated in \Cref{fig: leakage_example}, simply removing the function calls to \texttt{save} is insufficient. Residual information, such as import statements and surrounding logic (e.g., print statements), can still implicitly reveal the existence and semantics of the target function. In practice, such leakage can also persist in comments, docstrings, and the broader program structure. Therefore, simple rule-based filtering or static analysis alone is inadequate for eliminating all references to the target function. The desired indexing-time state is visualized in \Cref{fig: leakage_example}.

To construct a repository state that is both free of future context leakage and functionally intact, we introduce an automated agent, which we call the \emph{function removal agent}. This agent is equipped with tools for executing shell commands, reading and writing files, and navigating the repository. It is guided by static analysis tools \citep{tree_sitter} that identify all call sites of the target function, and it is prompted to explore and edit the repository to produce a realistic indexing-time state: one in which not only the callers but also all semantically linked information has been removed. Crucially, the agent is expected to preserve the functional correctness of the remaining codebase.

\begin{table*}[htbp]
    \centering
    \begin{adjustbox}{max width=\textwidth}
        \begin{tabular}{llrrr}
            \toprule
            Model & Method & Pass@1 (\%) & Latency (s) & Pre-processing Time (s) \\
            \midrule
            Qwen3-8B & None  & 16.22 & 4.03  & - \\
            Qwen3-8B & BM25  & 17.55 & 5.01  & - \\
            Qwen3-8B & RepoMap & 16.73 & 4.06  & 74.45 \\
            Qwen3-8B & Dense (UniXcoder) & 16.22 & 7.15  & - \\
            Qwen3-8B & Dense (CodeSage v2 large) & 15.41 & 11.50 & - \\
            Qwen3-8B & BM25 + RepoMap & 17.65 & 4.72  & 74.45 \\
            Qwen3-8B & Retriever Agent (Ours) & 20.92 & 4.36 & 46.27 \\
            Qwen3-8B & Forecaster Agent (Ours) & 22.35 & 4.55 & 49.08 \\
            Qwen3-8B & \textbf{SpecAgent (Ours)} & \textbf{27.86} & 4.55  & 49.08 \\
            \hdashline
            Qwen3-8B & \textit{Oracle Agent} & \textit{40.20} & 4.53  & 45.19 \\
            \midrule
            Qwen3-30B-A3B & None  & 18.37 & 5.03  & - \\
            Qwen3-30B-A3B & BM25  & 18.88 & 6.03  & - \\
            Qwen3-30B-A3B & RepoMap & 17.45 & 5.54  & 74.45 \\
            Qwen3-30B-A3B & Dense (UniXcoder) & 18.78 & 8.49  & - \\
            Qwen3-30B-A3B & Dense (CodeSage v2 large) & 16.94 & 14.84 & - \\
            Qwen3-30B-A3B & BM25 + RepoMap & 18.16 & 5.88  & 74.45 \\
            Qwen3-30B-A3B & Retriever Agent (Ours) & 23.27 & 5.75 & 46.27 \\
            Qwen3-30B-A3B & Forecaster Agent (Ours) & 23.67 & 5.80 & 49.08 \\
            Qwen3-30B-A3B & \textbf{SpecAgent (Ours)} & \textbf{27.96} & 5.89  & 49.08 \\
            \hdashline
            Qwen3-30B-A3B & \textit{Oracle Agent} & \textit{40.10} & 5.91  & 45.19 \\
            \bottomrule
        \end{tabular}
    \end{adjustbox}
    \caption{Pass@1, inference-time latency, and indexing pre-processing time for all retrieval methods.}
    \label{tab: exp_main}
\end{table*}

All experiments involving our retrieval agent are conducted on the indexing-time state produced by the function removal agent. This ensures a fair evaluation setting where no future knowledge of the target function is leaked. A detailed description of the construction pipeline, along with our benchmark validation procedure, is provided in \Cref{app: benchmark}. A real-world example of the synthetic indexing-time state constructed by the function removal agent is presented in \Cref{app: function_removal_example}. This example demonstrates that simple AST-based cleaning is insufficient to eliminate future context leakage.

\section{Experiments}\label{sec: exp}

\subsection{Experimental Setup}\label{sec: exp_setup}

We evaluate on the REPOCOD dataset \citep{repocod}, a function completion benchmark with 980 problems drawn from 11 popular open-source projects, where over 58\% of the problems require file- or repository-level context. For each problem, the completion model must generate the body of a target function given its signature and docstring, the left and right context, and optionally cross-file contexts.

We compare our indexing-time agents (including SpecAgent) and the oracle agent against the following baseline retrieval methods:  
(1) no cross-file context,
(2) sparse retrieval (BM25),
(3) RepoMap,
(4) BM25 + RepoMap, and
(5) dense retrieval using UniXcoder \citep{unixcoder} and CodeSage v2 large \citep{codesage}, the state-of-the-art low-latency retrievals for inline code completion.

A key distinction between baselines and our agents lies in the \emph{timing of context construction}:

\paragraph{Baseline retrieval methods.} Following the official REPOCOD setting, we remove the body of the target function at inference time, and use its signature and docstring as the query. Left and right file context as well as retrieved cross-file contexts are provided to the model. This setup enables comparability with prior work, but is known to suffer from \emph{future context leakage}, since other files may indirectly reference the ground-truth completion (\Cref{sec: benchmark_leakage}). As a result, baseline results likely \emph{overestimate} real-world deployment performance.

\paragraph{Indexing-time agents (ours).} Our agents operate on the synthetic indexing-time state of the repository (\Cref{sec: benchmark_creation}), before the target function has been written. They cannot rely on the ground-truth completion or future contexts, but instead must proactively predict or retrieve useful blocks based only on current repository contents. The resulting blocks are cached and later supplied to the completion model as cross-file context at inference. This setting avoids leakage and more accurately reflects realistic development scenarios, at the cost of being a harder task.

\paragraph{Oracle agent.} The Oracle Agent is evaluated at inference time, with access to the ground-truth completion. It generates cross-file context blocks that are passed to the completion model, but cannot include the ground-truth completion itself. This represents an upper bound.

\paragraph{Implementation details.} We evaluate with two code completion models: Qwen3-8B and Qwen3-30B-A3B \citep{qwen3}. All agents use Claude 3.7 Sonnet \citep{claude-3-7} as the backbone. We also experiment with Qwen3-Coder as the agent in \Cref{app: qwen3_coder}. The left, right, and cross-file contexts are each capped at 10K tokens. We report pass@1 as the primary metric, along with inference-time latency and indexing pre-processing time (\Cref{tab: exp_main}). We run experiments on a cluster with 8 A100 GPUs.

\subsection{Main Results}\label{sec: exp_main}

\Cref{tab: exp_main} shows that SpecAgent achieves the highest pass@1 among retrieval-based methods, excluding the Oracle Agent upper bound. SpecAgent's context improves the Qwen3-8B \citep{qwen3} model's performance by approximately \textbf{10--11\%} (58\% relative), and the Qwen3-30B-A3B \citep{qwen3} model by \textbf{9--10\%} (48\% relative) compared to strong baselines. Importantly, SpecAgent maintains inference-time efficiency: unlike BM25 or dense retrievers, it does not require on-the-fly index lookups or similarity computations, instead incurring a one-time indexing cost of roughly 50 seconds (executed asynchronously). A comparison of contexts generated by different methods is shown in \Cref{app: context_examples}.

\subsection{Ablation Studies}\label{sec: exp_ablation}

\paragraph{Component ablation.} We compare SpecAgent to its variants (Retriever Agent and Forecaster Agent, see \Cref{sec: method_indexing}). Removing either component lowers performance, highlighting the complementary roles of prediction and retrieval. Interestingly, the Forecaster Agent alone outperforms the Retriever Agent, underscoring the benefit of anticipating user intent.

\paragraph{Combined retrieval strategies.} When SpecAgent's contexts are concatenated with those from BM25 or dense retrievers, performance decreases (\Cref{tab: exp_combined}). This indicates that SpecAgent already selects high-quality contexts, and adding noisy blocks from other methods dilutes performance.

\begin{table}[htbp]
    \centering
    \begin{adjustbox}{max width=\columnwidth}
        \begin{tabular}{lr}
            \toprule
             Method & Pass@1 (\%) \\
             \midrule
             SpecAgent & \textbf{27.86} \\
             SpecAgent + BM25 & 25.10 \\
             SpecAgent + Dense (UniXcoder) & 25.31 \\
             \bottomrule
        \end{tabular}
    \end{adjustbox}
    \caption{Pass rates of combined retrieval strategies.}
    \label{tab: exp_combined}
\end{table}

\paragraph{Inference-time ablation.} For completeness, we also run SpecAgent under the same inference-time setup as baselines (target function removed, contexts retrieved at inference). This setting introduces leakage, enabling the agent to guess functionality from surrounding files. As shown in \Cref{tab: exp_inference_time}, SpecAgent achieves much higher pass@1 here, but we emphasize that these numbers are not realistic. Our main results adopt the stricter indexing-time async setting to better reflect real-world usage.

\begin{table}[htbp]
    \centering
    \begin{adjustbox}{max width=\columnwidth}
        \begin{tabular}{llr}
             \toprule
             Model & Method & Pass@1 (\%) \\
             \midrule
             Qwen3-8B & SpecAgent (indexing) & 27.86 \\
             Qwen3-8B & SpecAgent (inference) & \textbf{30.92} \\
             \midrule
             Qwen3-30B-A3B & SpecAgent (indexing) & 27.96 \\
             Qwen3-30B-A3B & SpecAgent (inference) & \textbf{34.29} \\
             \bottomrule
        \end{tabular}
    \end{adjustbox}
    \caption{Ablation on inference-time SpecAgent.}
    \label{tab: exp_inference_time}
\end{table}

\paragraph{Composition of context blocks.} We vary the ratio of prediction vs.\ retrieval blocks while fixing the total number. Performance peaks with three prediction blocks (\Cref{tab: exp_composition}), showing that both speculative predictions and retrieved contexts contribute to SpecAgent's success.

\begin{table}[htbp]
    \centering
    \begin{adjustbox}{max width=\columnwidth}
        \begin{tabular}{cc}
             \toprule
             \#Predictions & Pass@1 (\%) \\
             \midrule
             0 & 20.92 \\
             1 & 27.76 \\
             3 & \textbf{27.86} \\
             6 & 24.29 \\
             12 & 22.35 \\
             \bottomrule
        \end{tabular}
    \end{adjustbox}
    \caption{Ablation on the composition of contexts.}
    \label{tab: exp_composition}
\end{table}

\section{Summary}

We presented \emph{SpecAgent}, a speculative context construction framework that shifts repository-specific retrieval from inference time to indexing time, enabling large language models to operate with richer, pre-computed context while maintaining interactive responsiveness. By anticipating likely future changes and pre-gathering relevant cross-file information, SpecAgent addresses both the latency bottleneck and the context insufficiency that limit retrieval-augmented methods in real-world repositories. To enable realistic evaluation, we introduced a benchmark free from future-context leakage, providing a fairer measure of code completion performance. Experiments on two strong LLMs show consistent absolute 9--11\% accuracy gains over competitive baselines without additional inference-time cost, highlighting the promise of speculative context construction for scaling LLM-assisted software development to large, evolving codebases.

\section{Limitations}

The primary limitation of our work is the lack of publicly available benchmarks that eliminate \emph{future-context leakage}. To evaluate SpecAgent in a realistic setting, we constructed a synthetic benchmark by modifying REPOCOD with a function removal agent to create synthetic indexing-time repository states. While this setup removes leakage, it is not derived from actual real-world repository histories, and thus the resulting performance may differ from what would be observed in production environments. Conversely, experiments under the original REPOCOD setting yield high pass rates, but these results are inflated by future-context leakage and likewise fail to reflect real-world performance.

Beyond benchmark availability, our work also has several other limitations. First, SpecAgent's speculative capabilities depend on the quality and breadth of its indexing-time exploration tools; in repositories with unconventional structures or sparse documentation, relevant context may still be missed. Second, although we demonstrate latency benefits at inference time, SpecAgent introduces additional computational overhead at indexing time, which may be non-trivial for extremely large or frequently changing repositories. Finally, our experiments are limited to two strong code completion models (Qwen3-8B and Qwen3-30B-A3B); it remains to be seen how well the approach generalizes to smaller models, multilingual codebases, or tasks beyond function completion, such as bug fixing or large-scale refactoring.

\bibliography{references}

\clearpage
\onecolumn
\appendix
\renewcommand\thepart{}
\renewcommand\partname{}
\part{Appendix}
\setcounter{secnumdepth}{3}
\setcounter{tocdepth}{3}
\parttoc
\clearpage
\twocolumn

\section{Results with Qwen3-Coder as indexing-time agent}\label{app: qwen3_coder}

In \Cref{sec: exp}, we evaluated our indexing-time framework using Claude 3.7 Sonnet \citep{claude-3-7} as the agent and observed substantial gains over retrieval-based baselines. To assess generality across different agent backbones, we repeat the same experiments using Qwen3-Coder \citep{qwen3coder} as the indexing-time agent. Across all configurations, our agentic framework continues to outperform baseline retrieval methods, confirming its robustness and architecture-agnostic benefits.

\paragraph{Main results.} \Cref{tab: exp_main_qwen3_coder} reports the main results. Although Qwen3-Coder achieves slightly lower absolute scores than Claude 3.7 Sonnet (on average $\sim$5\% lower), the same trends hold: both the \emph{Retriever Agent} (retrieval only) and the \emph{Forecaster Agent} (prediction only) outperform all standard retrieval baselines, and the full \emph{SpecAgent}---combining retrieval and prediction---achieves the highest overall performance without introducing any inference-time latency. These results further validate that indexing-time synthesis and retrieval complement each other effectively.

\paragraph{Combined retrieval strategies.} As in \Cref{sec: exp_ablation}, we evaluate hybrid configurations combining SpecAgent's contexts with additional retrieved snippets from BM25 and dense retrievers. \Cref{tab: exp_combined_qwen3_coder} shows a modest improvement from adding dense retrieval (UniXcoder), indicating that SpecAgent's precomputed contexts already capture most relevant information.

\begin{table}[htbp]
    \centering
    \begin{adjustbox}{max width=\columnwidth}
        \begin{tabular}{lr}
            \toprule
             Method & Pass@1 (\%) \\
             \midrule
             SpecAgent & 20.20 \\
             SpecAgent + BM25 & 20.71 \\
             SpecAgent + Dense (UniXcoder) & \textbf{20.97} \\
             \bottomrule
        \end{tabular}
    \end{adjustbox}
    \caption{Effect of combining SpecAgent contexts with traditional retrievals. Qwen3-Coder is used as the indexing-time agent.}
    \label{tab: exp_combined_qwen3_coder}
\end{table}

\paragraph{Inference-time ablation.} We also test SpecAgent under an inference-time setup (retrieving contexts after removing the target function). This setting, while unrealistic due to information leakage, establishes an upper bound on achievable performance. As shown in \Cref{tab: exp_inference_time_qwen3_coder}, SpecAgent achieves higher pass@1 under this condition, consistent with trends reported in \Cref{sec: exp_ablation}.

\begin{table}[htbp]
    \centering
    \begin{adjustbox}{max width=\columnwidth}
        \begin{tabular}{llr}
             \toprule
             Model & Method & Pass@1 (\%) \\
             \midrule
             Qwen3-8B & SpecAgent (indexing) & 20.20 \\
             Qwen3-8B & SpecAgent (inference) & \textbf{24.08} \\
             \midrule
             Qwen3-30B-A3B & SpecAgent (indexing) & 24.69 \\
             Qwen3-30B-A3B & SpecAgent (inference) & \textbf{27.14} \\
             \bottomrule
        \end{tabular}
    \end{adjustbox}
    \caption{Inference-time ablation for SpecAgent using Qwen3-Coder. Higher results stem from leakage, not from improved generalization.}
    \label{tab: exp_inference_time_qwen3_coder}
\end{table}

\paragraph{Composition ablation.} Finally, we vary the ratio of retrieved versus predicted context blocks while fixing the total number (12). As shown in \Cref{tab: exp_composition_qwen3_coder}, performance peaks when using one or three prediction blocks---mirroring results from \Cref{sec: exp_ablation}. This reinforces that both speculative prediction and repository retrieval are essential for SpecAgent's success.

\begin{table}[htbp]
    \centering
    \begin{adjustbox}{max width=\columnwidth}
        \begin{tabular}{cc}
             \toprule
             \#Predictions & Pass@1 (\%) \\
             \midrule
             0 & 17.76 \\
             1 & \textbf{21.84} \\
             3 & 21.63 \\
             6 & 20.51 \\
             12 & 19.80 \\
             \bottomrule
        \end{tabular}
    \end{adjustbox}
    \caption{Ablation on the composition of SpecAgent's retrieved and predicted context blocks. Qwen3-Coder serves as the indexing-time agent.}
    \label{tab: exp_composition_qwen3_coder}
\end{table}

\begin{table*}[htbp]
    \centering
    \begin{adjustbox}{max width=\textwidth}
        \begin{tabular}{llrrr}
            \toprule
            Model & Method & Pass@1 (\%) & Latency (s) & Pre-processing Time (s) \\
            \midrule
            Qwen3-8B & None  & 16.22 & 4.03  & - \\
            Qwen3-8B & BM25  & 17.55 & 5.01  & - \\
            Qwen3-8B & RepoMap & 16.73 & 4.06  & 74.45 \\
            Qwen3-8B & Dense (UniXcoder) & 16.22 & 7.15  & - \\
            Qwen3-8B & Dense (CodeSage v2 large) & 15.41 & 11.50 & - \\
            Qwen3-8B & BM25 + RepoMap & 17.65 & 4.72  & 74.45 \\
            Qwen3-8B & Retriever Agent (Ours) & 17.76 & 4.26 & 41.45 \\
            Qwen3-8B & Forecaster Agent (Ours) & 19.80 & 3.95 & 40.72 \\
            Qwen3-8B & \textbf{SpecAgent (Ours)} & \textbf{20.20} & 3.80  & 41.45 \\
            \hdashline
            Qwen3-8B & \textit{Oracle Agent} & \textit{27.04} & 4.05  & 42.29 \\
            \midrule
            Qwen3-30B-A3B & None  & 18.37 & 5.03  & - \\
            Qwen3-30B-A3B & BM25  & 18.88 & 6.03  & - \\
            Qwen3-30B-A3B & RepoMap & 17.45 & 5.54  & 74.45 \\
            Qwen3-30B-A3B & Dense (UniXcoder) & 18.78 & 8.49  & - \\
            Qwen3-30B-A3B & Dense (CodeSage v2 large) & 16.94 & 14.84 & - \\
            Qwen3-30B-A3B & BM25 + RepoMap & 18.16 & 5.88  & 74.45 \\
            Qwen3-30B-A3B & Retriever Agent (Ours) & 20.20 & 5.28 & 41.45 \\
            Qwen3-30B-A3B & Forecaster Agent (Ours) & 21.73 & 5.72 & 40.72 \\
            Qwen3-30B-A3B & \textbf{SpecAgent (Ours)} & \textbf{24.69} & 5.39  & 41.45 \\
            \hdashline
            Qwen3-30B-A3B & \textit{Oracle Agent} & \textit{33.47} & 5.39  & 42.29 \\
            \bottomrule
        \end{tabular}
    \end{adjustbox}
    \caption{Main results using Qwen3-Coder as the indexing-time agent. SpecAgent achieves the highest pass@1 while maintaining inference-time latency comparable to baselines, demonstrating the benefits of combining retrieval and prediction at indexing time.}
    \label{tab: exp_main_qwen3_coder}
\end{table*}


\section{Agent implementation details}\label{app: agents}

This section documents the implementations of the agents introduced in \Cref{sec: method_indexing,sec: method_oracle,sec: benchmark_creation}. We focus on the operational details needed to reproduce our pipeline: tool access, execution workflow, and prompt templates. In the experiments reported in \Cref{sec: exp_setup}, we instantiate these agents with the backbone models described in the main text.

\subsection{Retriever agent}\label{app: agents_retriever}

The retriever agent is the retrieval-only indexing-time agent introduced in \Cref{sec: method_indexing}. It operates on the synthetic indexing-time repository state, after the target function has been removed, and produces a set of cross-file context blocks that are later cached and passed to the completion model at inference time.

\paragraph{Tools.} The agent is given two exploration tools: a file-reading tool for inspecting repository contents and a shell tool for lightweight repository navigation. Before retrieval begins, it invokes the function removal agent described in \Cref{app: agents_removal} to construct or reuse the leakage-free repository snapshot associated with the current REPOCOD instance.

\paragraph{Workflow.} For each benchmark instance, the system first locates a persistent function-removed repository snapshot. If such a snapshot is not already available, it copies the repository, removes the target function together with its callers, and stores the resulting snapshot for reuse. The retriever agent then reads the target file after removal, explores related files and import chains, and emits a fixed number of context blocks. Each block contains a short natural-language explanation and one focused piece of evidence, such as a code snippet, dependency pattern, validation idiom, or API usage example. These blocks are parsed into structured records and cached for later inference-time use.

\paragraph{Prompt.} The implementation uses a fixed system prompt together with an instance-specific task prompt. The output is parsed by looking for \texttt{<CONTEXT\_START>} and \texttt{<CONTEXT\_END>} delimiters.

\begin{minted}[fontsize=\scriptsize, breaklines, frame=single, bgcolor=gray!5, breakanywhere]{markdown}
[System prompt]
You are a code analysis agent that extracts relevant context for code completion tasks.

You have access to:
- The target file path: <TARGET_FILE_PATH>
- A repository where the target function has been removed
- All other files in the repository

Your task is to analyze the repository and extract contextual information that would help a code completion model when adding a new function to the target file. The target function has been cleanly removed from the repository, so you need to find relevant patterns, utilities, and examples from the remaining codebase.

RESPONSE FORMAT - Use this EXACT format:

<CONTEXT_START>
[Brief explanation of why this context is relevant]
[Your context content - can be code snippets, explanations, patterns, etc.]
<CONTEXT_END>

You can have multiple CONTEXT_START/CONTEXT_END blocks. Each block should contain ONE piece of relevant context.

ANALYSIS STRATEGY:
1. Start by examining the target file to understand its current structure and purpose
2. Analyze what kinds of features are missing from the target file, and what contexts would be helpful to add them
3. Look at import statements to understand dependencies
4. Find files in the same package/directory as the target file
5. Follow import chains to find relevant utility functions, classes, and patterns
6. Identify common patterns used throughout the codebase
7. Look for similar files or functions that might provide relevant context
8. Extract error handling, validation, and design patterns used in the codebase
9. Find configuration, constants, and data structures that might be relevant

CONTEXT TYPES YOU CAN PROVIDE:
- Code snippets from **other** files in the repository (with file paths)
- Import statements and dependency information
- Implementations and usages of imported functions and classes
- Class/function signatures and their usage patterns
- Error handling and validation patterns
- API usage examples
- Data structure definitions
- Constants and configuration values
- Test examples showing usage patterns
- Documentation and architectural insights
- Design patterns used in the codebase

GUIDELINES:
- Always include relative file paths when providing code snippets
- Do not include code snippets from the target file - they are already available to the code completion model
- Focus on context that would be helpful for someone adding a function to the target file
- Prioritize content from related files (same package, imported modules, etc.)
- Look for reusable patterns and utilities
- Each context should be focused and explain its relevance clearly
- Provide the most valuable context rather than quantity

TOOLS AVAILABLE:
- file_read: Read contents of specific files
- execute_bash: Run bash commands to explore repository structure

Remember: You are analyzing a repository where the target function has been cleanly removed. Your goal is to gather helpful context that would assist with adding a new function to the target file. The target file contents is already available to the code completion model, so please don't include them again in your contexts.

[Instance prompt]
TASK: Extract relevant context for adding a new function to a target file.

TARGET FILE: <TARGET_FILE_PATH>
REPOSITORY PATH: <TEMP_REPO_PATH>

TARGET FILE CONTENT (after function removal):
```python
<TARGET_FILE_CONTENT>
```

INSTRUCTIONS:
1. Analyze the repository structure to understand the codebase
2. Examine the target file to understand its current purpose and structure
3. Follow import statements to find relevant dependencies and utilities
4. Look for files in the same package/directory as the target file
5. Find similar patterns, utilities, and examples throughout the codebase
6. Focus on the <MAX_CONTEXT_FILES> most relevant files for context
7. Provide <MAX_CONTEXTS> context items using the required format

Since the target function has been removed, focus on:
- Understanding what the target file is for based on its current content
- Finding relevant utilities, patterns, and examples from the broader codebase
- Identifying common practices and conventions used in this repository
- Providing context that would be helpful for adding functionality to this type of file

Use the CONTEXT_START/CONTEXT_END format for each piece of context.
Provide a list of <MAX_CONTEXTS> context blocks in the CONTEXT_START/CONTEXT_END format, each containing one piece of relevant context.
Always include relative file paths when providing code snippets.
Use the file_read tool and execute_bash tool to explore the repository first before providing context.
\end{minted}

\subsection{Forecaster agent}\label{app: agents_forecaster}

The forecaster agent is the prediction-only indexing-time agent introduced in \Cref{sec: method_indexing}. Rather than retrieving repository evidence as the final output, it produces multiple plausible implementations of the missing function and uses these predicted implementations themselves as speculative context blocks.

\paragraph{Tools.} The forecaster agent uses the same repository exploration tools as the retriever agent: file reading and shell-based repository navigation. It also relies on the function removal agent to ensure that forecasting is performed on the synthetic indexing-time repository state rather than on the leakage-prone inference-time repository.

\paragraph{Workflow.} For each instance, the system first prepares or reuses the function-removed repository snapshot. The agent then inspects the target file after removal, studies repository structure and coding conventions, and proposes multiple candidate implementations for the missing function. Each candidate is intended to be a complete, standalone function body or function definition, representing a different plausible hypothesis about the future edit. These predicted implementations are stored as structured context blocks and cached for downstream composition in SpecAgent or for standalone ablations of the forecasting component.

\paragraph{Prompt.} As with the retriever agent, the implementation uses a fixed system prompt and an instance-specific task prompt. Predicted implementations are parsed from \texttt{<PREDICTION\_START>} and \texttt{<PREDICTION\_END>} blocks.

\begin{minted}[fontsize=\scriptsize, breaklines, frame=single, bgcolor=gray!5, breakanywhere]{markdown}
[System prompt]
You are a code prediction agent that generates multiple implementations of a missing function.

You have access to:
- The target file path: <TARGET_FILE_PATH>
- A repository where the target function has been removed
- All other files in the repository

Your task is to analyze the repository structure and generate MULTIPLE POSSIBLE IMPLEMENTATIONS of the missing function. You should NOT retrieve or quote existing code from other files - your job is purely to PREDICT what the missing function should look like.

RESPONSE FORMAT - Use this EXACT format:

<PREDICTION_START>
[Brief explanation of what this implementation does and why it might be correct]
```python
[Your predicted function implementation - complete Python code]
```
<PREDICTION_END>

You can have multiple PREDICTION_START/PREDICTION_END blocks. Each block should contain ONE complete function implementation.

PREDICTION STRATEGY:
1. Examine the target file to understand its structure, imports, and existing functionality
2. Analyze the file's context to understand what type of function might be missing
3. Look at the broader repository to understand coding patterns and conventions
4. Based on common patterns, generate multiple plausible implementations
5. Consider different complexity levels (simple, moderate, advanced implementations)
6. Think about edge cases and different approaches to the same problem

GUIDELINES:
- Generate ONLY function implementations - no other code snippets
- Each prediction should be a complete, standalone function
- Provide different variations/approaches for the same function
- Consider different levels of complexity and sophistication
- Include proper error handling, validation, and edge cases where appropriate
- Follow the coding style and patterns observed in the repository
- Make educated guesses based on function names, file context, and repository patterns
- DO NOT include code from other files - focus purely on prediction

TOOLS AVAILABLE:
- file_read: Read contents of specific files to understand context
- execute_bash: Run bash commands to explore repository structure

Remember: Your goal is to predict what the missing function should look like, not to retrieve existing code. Generate multiple creative and plausible implementations.

[Instance prompt]
TASK: Generate multiple predicted implementations for a missing function.

TARGET FILE: <TARGET_FILE_PATH>
REPOSITORY PATH: <TEMP_REPO_PATH>

TARGET FILE CONTENT (after function removal):
```python
<TARGET_FILE_CONTENT>
```

INSTRUCTIONS:
1. Analyze the target file structure and understand its purpose
2. Examine imports, class structure, and existing methods to understand context
3. Explore the repository structure to understand coding patterns and conventions
4. Generate <MAX_CONTEXTS> different predicted implementations for the missing function
5. Each prediction should be a complete, standalone function implementation
6. Consider different approaches: simple, moderate, and sophisticated implementations
7. Include proper error handling and edge cases where appropriate
8. Follow the coding style and patterns observed in the repository

FOCUS ON PREDICTION ONLY:
- DO NOT retrieve or quote existing code from other files
- Generate ONLY function implementations
- Each prediction should be creative and plausible
- Consider different complexity levels and approaches
- Think about what this function might realistically do based on its name and context

Use the PREDICTION_START/PREDICTION_END format for each predicted implementation.
Use the file_read tool and execute_bash tool to explore the repository first before providing context.
\end{minted}

\subsection{SpecAgent}\label{app: agents_spec}

SpecAgent is the hybrid indexing-time method studied throughout the paper. Conceptually, it combines the retriever agent and the forecaster agent described above. In the implementation, this combination is not realized as a separate interactive agent class. Instead, it is constructed deterministically by merging the cached outputs of the two component agents.

\paragraph{Tools.} SpecAgent inherits its effective tool access from its two constituent agents. During the composition stage itself, no additional interactive tools are required; the system only reads the cached outputs generated by the retriever and forecaster agents and writes the merged context list to a retrieval-compatible cache directory.

\paragraph{Workflow.} The implementation first runs the retriever agent and the forecaster agent independently over the evaluation set and caches their outputs. A composition script then loads both caches for each benchmark instance, keeps the first $n_{\mathrm{pred}}$ predicted implementations from the forecaster, and fills the remaining slots with the highest-ranked retrieval contexts from the retriever. The resulting ordered list is saved in the same format expected by the inference pipeline, so the completion model can consume SpecAgent contexts exactly as it would consume a standard agent cache. In the main experiments of \Cref{sec: exp_main}, we use three prediction blocks and nine retrieval blocks; \Cref{sec: exp_ablation} varies this ratio.

\paragraph{Prompt.} SpecAgent does not instantiate an additional LLM agent and therefore has no standalone prompt. The only prompt-bearing components are the retriever and forecaster agents in \Cref{app: agents_retriever,app: agents_forecaster}. The composition stage itself is deterministic: it loads the two caches for the same benchmark instance, keeps the first $n_{\mathrm{pred}}$ forecast blocks, appends retriever blocks until the context budget is filled, and saves the merged list in the cache format consumed by the inference pipeline.

\subsection{Oracle agent}\label{app: agents_oracle}

The oracle agent provides the upper-bound context construction setting discussed in \Cref{sec: method_oracle}. Unlike the indexing-time agents, it runs with access to the inference-time repository, the target function signature, and the full target-file contents. In the implementation, the oracle path uses the unmodified repository with full content, so the target-file contents supplied to the agent include the ground-truth implementation of the target function. The oracle is therefore an intentionally privileged upper bound rather than a realistic deployment setting.

\paragraph{Tools.} The oracle agent uses the same two repository exploration tools as the retriever agent: file reading and shell-based navigation. It does not call the function removal agent, because it is intentionally evaluated on the inference-time repository state.

\paragraph{Workflow.} For each benchmark instance, the system first reads the target file from the original repository and passes the resulting full file contents, together with the target function signature, to the oracle prompt. It also creates a temporary copy of the repository for exploration. The agent then searches the repository and emits structured context blocks. After generation, the parser applies a leakage filter before returning the final contexts. This filter removes blocks that appear to implement the target function by matching the target signature or strong implementation-oriented textual indicators in the content or explanation. The metadata records how many raw blocks were generated and how many survived filtering. This post-processing step is essential because the oracle agent is given direct access to the ground-truth target-file contents.

\paragraph{Prompt.} The implementation again uses a system prompt followed by an instance-specific task prompt. The text below matches the prompt strings in the codebase, with concrete paths and file contents replaced by placeholders.

\begin{minted}[fontsize=\scriptsize, breaklines, frame=single, bgcolor=gray!5, breakanywhere]{markdown}
[System prompt]
You are a specialized code analysis agent that extracts relevant context for code completion tasks.

Your task is to analyze a repository and extract contextual information that would help a code completion model complete a specific function. You must NOT provide the ground-truth implementation of the target function - only provide helpful context from OTHER files in the repository.

IMPORTANT: Use the following EXACT format for your responses:

<CONTEXT_START>
[Brief explanation of why this context is relevant]
[Your context content - can be code snippets, explanations, patterns, etc.]
<CONTEXT_END>

You can have multiple CONTEXT_START/CONTEXT_END blocks. Each block should contain ONE piece of relevant context.

CRITICAL CONSTRAINTS - CONTEXTS WILL BE AUTOMATICALLY REMOVED IF THEY VIOLATE THESE:
1. DO NOT provide any solution for the target function
2. DO NOT provide any code snippets from the target file: <TARGET_FILE_PATH>
3. ALWAYS include the relative file path when providing code snippets (e.g., "mypackage/utils/helpers.py")
4. Any context violating these rules will be automatically filtered out

ANALYSIS STRATEGY:
1. DO NOT provide any code snippets from the target file itself - that content is already available to the code completion model
2. The target file path is: <TARGET_FILE_PATH> - AVOID this file completely
3. Focus on analyzing OTHER files in the repository to find relevant context
4. Follow import statements to understand dependencies
5. Look for similar function patterns in related files (same package, similar names, etc.)
6. Identify utility functions, classes, and patterns that might be relevant
7. Extract common error handling and validation patterns used in this codebase
8. When providing code snippets, ALWAYS include the relative file path within the repository

CONTEXT TYPES YOU CAN PROVIDE (from OTHER files only):
- Code snippets from **other** files in the repository (with file paths)
- Import statements and dependency information from other files
- Class/function signatures and structures from related files
- Error handling patterns from other files
- Validation patterns from other files
- Similar function implementations from other files
- API usage patterns from other files
- Data structure definitions from other files
- Constants and configuration values from other files
- Test examples that show how similar functionality is used
- Documentation or comments that explain relevant concepts

ABSOLUTELY FORBIDDEN - THESE WILL BE AUTOMATICALLY REMOVED:
- Any completion of the target function
- Any code snippets from <TARGET_FILE_PATH>
- Any context without a clear relative file path for code snippets
- Any solutions or direct answers to what the target function should do
- Any code that looks like it could be the target function implementation

CONSTRAINTS:
- DO NOT provide any code snippets from the target file itself
- DO NOT provide solutions or implementations for the target function
- When providing code snippets, ALWAYS include the relative file path (e.g., "mypackage/utils/helpers.py")
- Prioritize quality over quantity - provide the most relevant context
- Each context block should be focused and specific
- Explain why each context is relevant in 1-2 sentences

TOOLS AVAILABLE:
- file_read: Read contents of specific files
- execute_bash: Run bash commands to explore repository structure

Remember: Your job is to provide helpful CONTEXT from OTHER files, not to solve the problem. The context will be provided as hints to help a code completion model. Any contexts that complete the target function, provide code from the target file, or lack proper file paths will be automatically removed.

[Instance prompt]
TASK: Extract relevant context for completing a Python function.

TARGET FILE: <TARGET_FILE_PATH>
REPOSITORY PATH: <TEMP_REPO_PATH>

FUNCTION TO COMPLETE:
<FUNCTION_SIGNATURE>

TARGET FILE CONTENT:
```python
<TARGET_FILE_CONTENT>
```

INSTRUCTIONS:
1. Explore the repository structure to understand the codebase
2. DO NOT provide any code snippets from the target file (<TARGET_FILE_PATH>) - it is already available to the code completion model
3. Find similar functions, utility classes, and relevant patterns in OTHER files
4. Extract context that would help complete the target function - but DO NOT implement the target function
5. Focus on the <MAX_CONTEXT_FILES> most relevant files (excluding the target file)
6. Provide up to <MAX_CONTEXTS> context items using the required format

CRITICAL WARNINGS - These contexts will be automatically REMOVED:
- Any implementation or completion of the target function
- Any code snippets from <TARGET_FILE_PATH>
- Any code snippets without relative file paths
- Any solutions that show how to implement the target function

Remember to use the CONTEXT_START/CONTEXT_END format for each piece of context you provide.
Do NOT provide the ground truth implementation of the target function - only provide helpful context from OTHER files.
Do NOT provide any code snippets from the target file itself.
Include relative file paths when providing code snippets from other files.
Focus on patterns, utilities, and examples from OTHER files that could inform the implementation.
Use the file_read tool and execute_bash tool to explore the repository first before providing context.
\end{minted}

\subsection{Function removal agent}\label{app: agents_removal}

The function removal agent constructs the synthetic indexing-time repository state used by the retriever and forecaster agents. Its purpose is to remove the target function and all semantically revealing callers while preserving syntactic validity and, as far as possible, the functional coherence of the remaining repository. This agent is therefore the core mechanism that eliminates future context leakage in the benchmark described in \Cref{sec: benchmark_creation,app: benchmark}.

\paragraph{Tools.} This agent has a richer tool set than the context-construction agents. In addition to file reading and shell navigation, it is given repository editing tools together with tree-sitter-based function and class removal tools. The tree-sitter analysis identifies candidate references to the target function, including top-level functions, class methods, decorated functions, and qualified calls, and the agent then uses the editing tools to carry out repository-wide removal and cleanup.

\paragraph{Workflow.} Given a repository snapshot and a target function, the system first enumerates all Python files and uses tree-sitter to detect candidate call sites and, when needed, the target function definition itself. It then summarizes these references for the LLM agent and asks the agent to remove the function definition, delete or rewrite all callers, clean up now-unused imports, and preserve syntactic validity. The resulting repository snapshot is stored persistently and reused across later retrieval and forecasting runs. In practice, this procedure removes not only direct calls but also many indirect clues, such as imports and local scaffolding that would otherwise reveal the target function's intended behavior.

\paragraph{Prompt.} The removal pipeline again uses a reusable system prompt and an instance-specific task prompt. The second prompt contains a tree-sitter-derived summary of candidate references before the LLM edits the repository.

\begin{minted}[fontsize=\scriptsize, breaklines, frame=single, bgcolor=gray!5, breakanywhere]{markdown}
[System prompt]
You are a specialized code modification agent that removes specific functions and all their callers from Python files.

Your task is to analyze Python files and remove:
1. The target function definition
2. All calls to the target function throughout the repository
3. Any import statements that become unused after removal

While preserving:
1. File structure and syntax
2. Other functions and classes
3. Necessary import statements
4. Module-level code
5. Comments and docstrings (except those within removed code)

RESPONSE FORMAT - Use this EXACT format:

<RESULT>
[Brief explanation of what was done]
<RESULT_END>

ANALYSIS STRATEGY:
1. Read and understand the code structure
2. Identify the target function and all its call sites
3. Remove function calls intelligently:
   - If the call is a standalone statement, remove the entire statement
   - If the call is part of an expression, remove or replace appropriately
   - If the call is in a condition, handle the conditional logic
   - If the call is in an assignment, remove or modify the assignment
4. Remove the function definition itself
5. Clean up any unused imports
6. Ensure all modified files remain syntactically valid
7. Report comprehensive details of what was removed

TOOLS AVAILABLE:
- file_read: Read contents of specific files
- execute_bash: Run bash commands to explore repository structure
- editor: Edit files (view, str_replace, etc.)
- remove_class: Remove entire classes from Python files using tree-sitter
- remove_function: Remove specific functions from Python files using tree-sitter

CRITICAL CONSTRAINTS:
- Remove the target function AND all its callers
- Do not leave any comments about the removal
- Preserve proper indentation and formatting
- Ensure all files remain syntactically valid
- Handle edge cases like method calls, decorated functions, etc.
- Be intelligent about removing calls without breaking code flow
- Report if the target function cannot be found

Remember: Your goal is complete removal of the target function and all references to it while keeping the repository functional.

[Instance prompt]
TASK: Remove a specific function and ALL its callers from a Python repository.

REPOSITORY PATH: <TEMP_REPO_PATH>
TARGET FILE: <TARGET_FILE_PATH>
FUNCTION TO REMOVE: <FUNCTION_NAME>

FUNCTION REFERENCES FOUND:
<REFERENCES_SUMMARY>

INSTRUCTIONS:
1. Remove the function definition from <TARGET_FILE_PATH>
2. Remove ALL calls to this function from the repository
3. For each function call, handle removal intelligently:
   - If it's a standalone statement: Remove the entire statement
   - If it's part of an assignment: Remove or modify the assignment appropriately
   - If it's in a conditional: Handle the condition logic properly
   - If it's in an expression: Replace or remove the expression safely
   - If it's a method call: Remove the call and handle the result appropriately

4. Ensure all modified files remain syntactically valid
5. Clean up any imports that become unused after removing calls
6. Handle edge cases like:
   - Function calls in list comprehensions
   - Function calls as arguments to other functions
   - Function calls in return statements
   - Method calls on objects

CRITICAL REQUIREMENTS:
- The function <FUNCTION_NAME> should not exist ANYWHERE in the repository after removal
- Do not leave any comments about the removal
- All Python files should remain syntactically valid
- Preserve the overall structure and functionality of unrelated code
- Be intelligent about removing calls without breaking program flow

FILES TO POTENTIALLY MODIFY:
<PYTHON_FILE_LIST>

Please proceed systematically through each file and remove all references to <FUNCTION_NAME>.
Report comprehensively what was removed from each file.
\end{minted}

\subsection{Function removal scoring agent}\label{app: agents_scoring}

The function removal scoring agent is used in the benchmark validation loop summarized in \Cref{app: benchmark}. Its purpose is to assess whether the synthetic indexing-time repository state is sufficiently leakage-free to be retained in the benchmark. The agent does not modify the repository. Instead, it inspects the post-removal snapshot and assigns a structured quality score.

\paragraph{Tools.} The scoring agent uses only repository exploration tools: file reading and shell-based navigation. It receives the target file path and target function name and is asked to reason about whether any remaining definitions or suspicious calls are likely to refer to the removed function.

\paragraph{Workflow.} The validator first runs basic syntactic checks using tree-sitter and then invokes the scoring agent on the repository snapshot. The agent examines the target file for residual definitions, searches for potentially related calls across the repository, reasons about import structure and call context, and returns a score in the range 0--5 together with a confidence value and a free-form explanation. Repositories with scores below a threshold of 4 are treated as problematic and are either refined or regenerated, yielding the iterative validation loop described in \Cref{app: benchmark}.

\paragraph{Prompt.} The scoring agent also uses a fixed system prompt plus an instance-specific task prompt. The system prompt defines the response fields and the 0--5 scale used by the validator.

\begin{minted}[fontsize=\scriptsize, breaklines, frame=single, bgcolor=gray!5, breakanywhere]{markdown}
[System prompt]
You are a specialized code analysis agent that evaluates how well a specific function has been removed from a Python repository.

Your task is to analyze a Python repository and provide a detailed assessment of function removal quality.

ANALYSIS APPROACH:
1. Check if the target function definition still exists (critical factor)
2. Identify function calls that might be related to the target function
3. Use context clues to determine if remaining calls are likely to the target function:
   - Import patterns
   - Variable assignment patterns
   - Class instantiation patterns
   - Method chaining patterns
   - Documentation and comments

4. Assess the likelihood that remaining calls are actually calls to the target function
5. Consider edge cases like inheritance, polymorphism, and name collisions

SCORING CRITERIA (0-5 scale):
- 5: Perfect removal - function definition gone, no suspicious calls remain
- 4: Good removal - definition gone, only clearly unrelated calls with same name remain
- 3: Moderate removal - definition gone, some potentially related calls remain
- 2: Poor removal - definition gone, many likely related calls remain
- 1: Very poor removal - definition exists OR many obvious calls to target function remain
- 0: No removal detected - target function definition clearly exists and unchanged

RESPONSE FORMAT - Use this EXACT format:

<ANALYSIS>
[Detailed analysis of what you found]
<ANALYSIS_END>

<SCORE>
[Integer score 0-5]
<SCORE_END>

<CONFIDENCE>
[Float confidence 0.0-1.0]
<CONFIDENCE_END>

<EXPLANATION>
[Human-readable explanation of the score]
<EXPLANATION_END>

TOOLS AVAILABLE:
- file_read: Read contents of specific files
- execute_bash: Run bash commands to explore repository structure

CRITICAL CONSTRAINTS:
- Focus on the SPECIFIC target function, not all functions with the same name
- Consider context and import patterns to distinguish function calls
- Be conservative - if unsure whether a call is to the target function, assume it might be
- Provide detailed reasoning for your score
- Consider both false positives (unrelated functions) and false negatives (missed calls)

Remember: The goal is accurate assessment of removal quality for the SPECIFIC target function.

[Instance prompt]
TASK: Analyze and score how well a specific function has been removed from a Python repository.

REPOSITORY PATH: <TEMP_REPO_PATH>
TARGET FILE: <TARGET_FILE_PATH>
FUNCTION TO ASSESS: <FUNCTION_NAME>

ANALYSIS INSTRUCTIONS:
1. First, check if the target function definition still exists in <TARGET_FILE_PATH>
   - Look for the exact function: <FUNCTION_NAME>
   - Consider decorated functions, class methods, etc.

2. Search the entire repository for potential calls to this function
   - Look for direct calls: <FUNCTION_NAME>()
   - Look for qualified calls: obj.<SIMPLE_NAME>()
   - Look for import-based calls

3. For each potential call, assess the likelihood it's actually calling the target function:
   - Check import statements and module paths
   - Look at variable assignments and object instantiation
   - Consider class inheritance and method resolution
   - Analyze context clues in surrounding code

4. Pay special attention to:
   - Class methods vs. instance methods
   - Functions with common names that might exist in multiple classes
   - Method chaining and attribute access patterns
   - Test files that might import or reference the function

5. Provide a score (0-5) based on removal quality:
   - How thoroughly was the target function removed?
   - How many likely calls to the target function remain?
   - What's the confidence that remaining calls are NOT to the target function?

Please analyze the repository systematically and provide your assessment.
\end{minted}


\section{Benchmark construction details}\label{app: benchmark}

This appendix provides additional details on the construction of our benchmark for indexing-time context retrieval. We describe both the creation of indexing-time repository states and the validation process used to ensure their quality.

\subsection{Benchmark creation}

Given a target function and its associated inference-time repository, our goal is to synthesize a plausible state of the repository from an earlier point in time, before the target function has been implemented. This reflects a realistic scenario in which the user has not yet begun authoring the target function, and the retrieval agent operates asynchronously during indexing time. As discussed in \Cref{sec: benchmark_leakage}, residual references to the target function can remain even after removing its body, including call sites, imports, and docstrings.

To construct a repository state that is both free of future context leakage and functionally intact, we introduce a \emph{function removal agent}. This agent is guided by static analysis tools and is equipped with shell and file manipulation tools to explore the repository. It edits files to eliminate all explicit and implicit references to the target function while preserving the functional correctness of the remaining codebase. The indexing-time state produced by this process serves as the foundation for all experiments in the main paper.

\subsection{Benchmark validation}

To ensure the quality and reliability of our synthetic benchmark, we implement a validation procedure that evaluates and refines the constructed indexing-time repository states. Specifically, we introduce a \emph{function removal scoring agent} that explores the repository and assigns a score to the quality of the function removal.

\begin{figure}[htbp]
    \centering
    \begin{tikzpicture}[
        node distance=1.5cm and 1.8cm,
        every node/.style={font=\small},
        box/.style={draw, rounded corners, minimum width=1.5cm, minimum height=1cm, align=center},
        arr/.style={-{Stealth[length=3mm]}, thick},
        square/.style={draw, minimum width=1.5cm, minimum height=1cm, align=center}
    ]
        \node[box, fill=green!20] (repo) {repo};
        \node[square, fill=orange!20, right=of repo] (removal) {removal\\agent};
        \node[box, fill=cyan!20, below=of removal] (processed) {indexing-time\\repo};
        \node[square, fill=magenta!20, below=of repo] (scoring) {scoring\\agent};

        \draw[arr] (repo) -- (removal);
        \draw[arr] (removal) -- (processed);
        \draw[arr] (processed) -- (scoring);
        \draw[arr] (scoring) -- node[left, align=center, xshift=-0.1cm]{score $<4$} (removal);
    \end{tikzpicture}
    \caption{The benchmark validation loop.}
    \label{fig: benchmark_validation}
\end{figure}

The scoring agent is prompted to analyze whether all references to the target function have been removed and whether the repository remains functional. It produces a quality score from 0 to 5. If the score falls below 4, the function removal agent is reapplied to further refine the repository state. This loop is repeated until the repository receives a score of at least 4. We apply this iterative validation and refinement pipeline across all target functions in our dataset to ensure that future context leakage is effectively mitigated and that the benchmark reliably reflects indexing-time retrieval scenarios.


\section{Function removal example}\label{app: function_removal_example}

In the main text, we introduced the issue of \emph{future-context leakage} in existing code completion benchmarks and proposed a method to construct synthetic indexing-time repository states that avoid leaking information about the target function. We argued that relying solely on static analysis tools to remove callers of the target function is insufficient, since semantic and contextual clues may still enable an agent to infer the target function. To address this, we introduced a \emph{function removal agent} that constructs the synthetic indexing-time state, as well as a \emph{function removal scoring agent} that validates the quality of this process.

In this section, we provide a concrete example from the seaborn~\citep{seaborn} repository. We illustrate the states of both the target file (containing the target function) and a caller file (which imports and invokes the target function) under inference-time and indexing-time states. We also highlight what information is removed in the synthetic indexing-time state and explain why static analysis alone cannot achieve the same effect.

\subsection{Inference-time state}\label{app: inference_time_state}

We examine an example from the REPOCOD benchmark~\citep{repocod}, specifically repository ID 33 corresponding to the seaborn repository. The target function is \texttt{color\_palette}, defined in \texttt{seaborn/palettes.py}. The inference-time state of the target file is shown below, with the definition and reference to the target function highlighted.

\begin{minted}[fontsize=\scriptsize, breaklines, frame=single, escapeinside=||, bgcolor=gray!5, breakanywhere]{python}
import colorsys
from itertools import cycle
import numpy as np
import matplotlib as mpl
from .external import husl
from .utils import desaturate, get_color_cycle
from .colors import xkcd_rgb, crayons
from ._compat import get_colormap


__all__ = [|\bfseries{|"color_palette"|}|, "hls_palette", "husl_palette", "mpl_palette",
           "dark_palette", "light_palette", "diverging_palette",
           "blend_palette", "xkcd_palette", "crayon_palette",
           "cubehelix_palette", "set_color_codes"]

...

class _ColorPalette(list):
    ...

def _patch_colormap_display():
    ...

|\bfseries{|def color_palette(palette=None, n_colors=None, desat=None, as_cmap=False):|}|
    """Return a list of colors or continuous colormap defining a palette.

    ..."""
    if palette is None:
        palette = get_color_cycle()
        if n_colors is None:
            n_colors = len(palette)

    elif not isinstance(palette, str):
        palette = palette
        if n_colors is None:
            n_colors = len(palette)
    else:
        ...

    return palette

def hls_palette(n_colors=6, h=.01, l=.6, s=.65, as_cmap=False):  # noqa
    ...
\end{minted}

As seen in the code, the target function \texttt{color\_palette} is defined just before the \texttt{hls\_palette} function and is also referenced in the \texttt{\_\_all\_\_} variable at the beginning of the file. The latter reference introduces future context leakage, since retrieval methods may leverage the \texttt{\_\_all\_\_} declaration to deduce the existence of the target function.

We also examine another file, \texttt{seaborn/axisgrid.py}, which imports and calls the target function. The inference-time state of this caller file is shown below, with references to the target function highlighted.

\begin{minted}[fontsize=\scriptsize, breaklines, frame=single, escapeinside=||, bgcolor=gray!5, breakanywhere]{python}
from __future__ import annotations
from itertools import product
from inspect import signature
import warnings
from textwrap import dedent

...
from .palettes import |\bfseries{|color_palette|}|, blend_palette
from ._docstrings import (
    DocstringComponents,
    _core_docs,
)

__all__ = ["FacetGrid", "PairGrid", "JointGrid", "pairplot", "jointplot"]

_param_docs = DocstringComponents.from_nested_components(
    core=_core_docs["params"],
)

...

class Grid(_BaseGrid):
    """A grid that can have multiple subplots and an external legend."""
    _margin_titles = False
    _legend_out = True

    def __init__(self):
        ...

    def _get_palette(self, data, hue, hue_order, palette):
        """Get a list of colors for the hue variable."""
        if hue is None:
            palette = |\bfseries{|color_palette|}|(n_colors=1)

        else:
            hue_names = categorical_order(data[hue], hue_order)
            n_colors = len(hue_names)

            # By default use either the current color palette or HUSL
            if palette is None:
                current_palette = utils.get_color_cycle()
                if n_colors > len(current_palette):
                    colors = |\bfseries{|color_palette|}|("husl", n_colors)
                else:
                    colors = |\bfseries{|color_palette|}|(n_colors=n_colors)

            # Allow for palette to map from hue variable names
            elif isinstance(palette, dict):
                color_names = [palette[h] for h in hue_names]
                colors = |\bfseries{|color_palette|}|(color_names, n_colors)

            # Otherwise act as if we just got a list of colors
            else:
                colors = |\bfseries{|color_palette|}|(palette, n_colors)

            palette = |\bfseries{|color_palette|}|(colors, n_colors)

        return palette
\end{minted}

Here, the target function is imported at the top of the file and called multiple times inside the \texttt{\_get\_palette} function. This poses a strong leakage risk: for example, BM25 may retrieve code chunks containing these call sites, which would not exist in real-world settings where the user has not yet implemented the function. By analyzing such call sites, an indexing-time agent could infer the intended behavior of the missing function and even generate its complete implementation, artificially inflating benchmark performance.

Attempting to construct synthetic indexing-time states with static analysis tools---for example, by simply removing the lines that call the target function---proves inadequate. Such removal leaves behind empty or broken control structures, from which an intelligent agent can still deduce the intended role of the function. This motivates our use of an agent-based approach to reliably construct leakage-free states, as demonstrated below.

\subsection{Indexing-time state}\label{app: indexing-time-state}

We now show the indexing-time states of the target and caller files produced by the function removal agent. For the target file, both the function definition and its reference in the \texttt{\_\_all\_\_} variable are removed, eliminating obvious leakage channels.
\begin{minted}[fontsize=\scriptsize, breaklines, frame=single, escapeinside=||, bgcolor=gray!5, breakanywhere]{python}
import colorsys
from itertools import cycle

import numpy as np
import matplotlib as mpl

from .external import husl

from .utils import desaturate, get_color_cycle
from .colors import xkcd_rgb, crayons
from ._compat import get_colormap


__all__ = ["hls_palette", "husl_palette", "mpl_palette",
           "dark_palette", "light_palette", "diverging_palette",
           "blend_palette", "xkcd_palette", "crayon_palette",
           "cubehelix_palette", "set_color_codes"]

...

class _ColorPalette(list):
    ...


def _patch_colormap_display():
    ...


def hls_palette(n_colors=6, h=.01, l=.6, s=.65, as_cmap=False):  # noqa
    ...
\end{minted}

For the caller file, the function removal agent eliminates both the import statement and all call sites of the target function. To preserve code functionality and avoid leaving broken logic, the agent further introduces a helper function, \texttt{\_process\_palette}, as a replacement for the removed function. This ensures that the caller file remains coherent and executable, while also preventing any information leakage that could enable an indexing-time agent or retrieval method to infer the target function. In this way, the synthetic indexing-time state avoids future context leakage and provides a more realistic evaluation environment.
\begin{minted}[fontsize=\scriptsize, breaklines, frame=single, escapeinside=||, bgcolor=gray!5, breakanywhere]{python}
from __future__ import annotations
from itertools import product, cycle
from inspect import signature
import warnings
from textwrap import dedent

...
from .palettes import blend_palette, husl_palette, SEABORN_PALETTES
from ._docstrings import (
    DocstringComponents,
    _core_docs,
)

|\bfseries{|def _process_palette(palette=None, n_colors=None):|}|
    """Internal palette processing function."""
    if palette is None:
        palette = get_color_cycle()
        if n_colors is None:
            n_colors = len(palette)
    elif not isinstance(palette, str):
        if n_colors is None:
            n_colors = len(palette)
    else:
        ...
    return palette

__all__ = ["FacetGrid", "PairGrid", "JointGrid", "pairplot", "jointplot"]

_param_docs = DocstringComponents.from_nested_components(
    core=_core_docs["params"],
)

...

class Grid(_BaseGrid):
    """A grid that can have multiple subplots and an external legend."""
    _margin_titles = False
    _legend_out = True

    def __init__(self):
        ...

    def _get_palette(self, data, hue, hue_order, palette):
        """Get a list of colors for the hue variable."""
        if hue is None:
            palette = |\bfseries{|_process_palette|}|(n_colors=1)

        else:
            hue_names = categorical_order(data[hue], hue_order)
            n_colors = len(hue_names)

            # By default use either the current color palette or HUSL
            if palette is None:
                current_palette = utils.get_color_cycle()
                if n_colors > len(current_palette):
                    colors = husl_palette(n_colors)
                else:
                    colors = |\bfseries{|_process_palette|}|(n_colors=n_colors)

            # Allow for palette to map from hue variable names
            elif isinstance(palette, dict):
                color_names = [palette[h] for h in hue_names]
                colors = |\bfseries{|_process_palette|}|(color_names, n_colors)

            # Otherwise act as if we just got a list of colors
            else:
                colors = |\bfseries{|_process_palette|}|(palette, n_colors)

            palette = |\bfseries{|_process_palette|}|(colors, n_colors)

        return palette
\end{minted}


\section{Code completion prompt example}\label{app: prompt_example}

The construction of prompts for the code completion model varies depending on the retrieval method. Each prompt is designed to provide the model with sufficient contextual information to complete the target function accurately. In particular, the prompt is structured as follows:
\begin{enumerate}
    \item \emph{Target file path}: The path to the source file containing the target function.
    \item \emph{Left context}: The content preceding the target function within the same file.
    \item \emph{Right context}: The content following the target function within the same file.
    \item \emph{Cross-file contexts} (optional): Additional relevant contexts retrieved from other files in the repository.
    \item \emph{Function signature and docstring}: The header and documentation string of the target function.
\end{enumerate}

To illustrate, we provide below an example prompt constructed using the BM25 retrieval method. The example is drawn from the seaborn~\citep{seaborn} repository, corresponding to repository ID 0 in the REPOCOD benchmark. The full prompt presented to the code completion model is as follows:
\begin{minted}[fontsize=\scriptsize, breaklines, frame=single,  bgcolor=gray!5, breakanywhere]{markdown}
This is the file that contains the target function to be generated.

## File path: seaborn/_core/scales.py

### Context before the target function
```python
from __future__ import annotations
import re
from copy import copy
from collections.abc import Sequence
from dataclasses import dataclass
from functools import partial
from typing import Any, Callable, Tuple, Optional, ClassVar
...
from matplotlib.axis import Axis
from matplotlib.scale import ScaleBase
from pandas import Series

from seaborn._core.rules import categorical_order
from seaborn._core.typing import Default, default

from typing import TYPE_CHECKING
...


@dataclass
class Continuous(ContinuousBase):
    """
    A numeric scale supporting norms and functional transforms.
    """
    values: tuple | str | None = None
    trans: str | TransFuncs | None = None

    # TODO Add this to deal with outliers?
    # outside: Literal["keep", "drop", "clip"] = "keep"

    _priority: ClassVar[int] = 1

    def tick(
        self,
        locator: Locator | None = None, *,
        at: Sequence[float] | None = None,
        upto: int | None = None,
        count: int | None = None,
        every: float | None = None,
        between: tuple[float, float] | None = None,
        minor: int | None = None,
    ) -> Continuous:
        """
        Configure the selection of ticks for the scale's axis or legend.

        ..."""
        # Input checks
        if locator is not None and not isinstance(locator, Locator):
            raise TypeError(
                f"Tick locator must be an instance of {Locator!r}, "
                f"not {type(locator)!r}."
            )
        log_base, symlog_thresh = self._parse_for_log_params(self.trans)
        if log_base or symlog_thresh:
            if count is not None and between is None:
                raise RuntimeError("`count` requires `between` with log transform.")
            if every is not None:
                raise RuntimeError("`every` not supported with log transform.")

        new = copy(self)
        new._tick_params = {
            "locator": locator,
            "at": at,
            "upto": upto,
            "count": count,
            "every": every,
            "between": between,
            "minor": minor,
        }
        return new
```

### Context after the target function
```python
    def _parse_for_log_params(
        self, trans: str | TransFuncs | None
    ) -> tuple[float | None, float | None]:
        ...

    def _get_locators(self, locator, at, upto, count, every, between, minor):
        ...

    def _get_formatter(self, locator, formatter, like, base, unit):
        ...


@dataclass
class Temporal(ContinuousBase):
    """
    A scale for date/time data.
    """
    ...

...
```

### Relevant context from other files of the repo
```python
# Code from: seaborn/categorical.py
    width=dedent("""\
    width : float
        Width allotted to each element on the orient axis. When `native_scale=True`,
        it is relative to the minimum distance between two values in the native scale.\
    """),
    dodge=dedent("""\
    dodge : "auto" or bool
        When hue mapping is used, whether elements should be narrowed and shifted along
        the orient axis to eliminate overlap. If `"auto"`, set to `True` when the
        orient variable is crossed with the categorical variable or `False` otherwise.
        .. versionchanged:: 0.13.0
            Added `"auto"` mode as a new default.\
    """),
    linewidth=dedent("""\
    linewidth : float
        Width of the lines that frame the plot elements.\
    """),
    linecolor=dedent("""\
    linecolor : color
        Color to use for line elements, when `fill` is True.
        .. versionadded:: v0.13.0\
    """),
    log_scale=dedent("""\
    log_scale : bool or number, or pair of bools or numbers
        Set axis scale(s) to log. A single value sets the data axis for any numeric
        axes in the plot. A pair of values sets each axis independently.
        Numeric values are interpreted as the desired base (default 10).
        When `None` or `False`, seaborn defers to the existing Axes scale.
        .. versionadded:: v0.13.0\
    """),
    native_scale=dedent("""\
    native_scale : bool
        When True, numeric or datetime values on the categorical axis will maintain
        their original scaling rather than being converted to fixed indices.
        .. versionadded:: v0.13.0\
    """),
    formatter=dedent("""\
    formatter : callable
        Function for converting categorical data into strings. Affects both grouping
        and tick labels.
        .. versionadded:: v0.13.0\
    """),
    legend=dedent("""\
    legend : "auto", "brief", "full", or False
        How to draw the legend. If "brief", numeric `hue` and `size`
        variables will be represented with a sample of evenly spaced values.
        If "full", every group will get an entry in the legend. If "auto",
        choose between brief or full representation based on number of levels.
        If `False`, no legend data is added and no legend is drawn.
        .. versionadded:: v0.13.0\
    """),

# Code from: seaborn/_core/plot.py
"variables": variables,
            "structure": structure,
            "wrap": wrap,
        }
        new = self._clone()
        new._facet_spec.update(spec)
        return new
    # TODO def twin()?
    def scale(self, **scales: Scale) -> Plot:
        """
        Specify mappings from data units to visual properties.
        Keywords correspond to variables defined in the plot, including coordinate
        variables (`x`, `y`) and semantic variables (`color`, `pointsize`, etc.).
        A number of "magic" arguments are accepted, including:
            - The name of a transform (e.g., `"log"`, `"sqrt"`)
            - The name of a palette (e.g., `"viridis"`, `"muted"`)
            - A tuple of values, defining the output range (e.g. `(1, 5)`)
            - A dict, implying a :class:`Nominal` scale (e.g. `{"a": .2, "b": .5}`)
            - A list of values, implying a :class:`Nominal` scale (e.g. `["b", "r"]`)
        For more explicit control, pass a scale spec object such as :class:`Continuous`
        or :class:`Nominal`. Or pass `None` to use an "identity" scale, which treats
        data values as literally encoding visual properties.
        Examples
        --------
        .. include:: ../docstrings/objects.Plot.scale.rst
        """
        new = self._clone()
        new._scales.update(scales)
        return new
    def share(self, **shares: bool | str) -> Plot:
        """
        Control sharing of axis limits and ticks across subplots.
        Keywords correspond to variables defined in the plot, and values can be
        boolean (to share across all subplots), or one of "row" or "col" (to share
        more selectively across one dimension of a grid).
        Behavior for non-coordinate variables is currently undefined.
        Examples
        --------
        .. include:: ../docstrings/objects.Plot.share.rst
        """
        new = self._clone()
        new._shares.update(shares)
        return new
    def limit(self, **limits: tuple[Any, Any]) -> Plot:
        """
        Control the range of visible data.
        Keywords correspond to variables defined in the plot, and values are a
        `(min, max)` tuple (where either can be `None` to leave unset).
        Limits apply only to the axis; data outside the visible range are
        still used for any stat transforms and added to the plot.
        """

...
```

### Target function to complete

```python
    def label(
        self,
        formatter: Formatter | None = None, *,
        like: str | Callable | None = None,
        base: int | None | Default = default,
        unit: str | None = None,
    ) -> Continuous:
        """
        Configure the appearance of tick labels for the scale's axis or legend.

        Parameters
        ----------
        formatter : :class:`matplotlib.ticker.Formatter` subclass
            Pre-configured formatter to use; other parameters will be ignored.
        like : str or callable
            Either a format pattern (e.g., `".2f"`), a format string with fields named
            `x` and/or `pos` (e.g., `"${x:.2f}"`), or a callable with a signature like
            `f(x: float, pos: int) -> str`. In the latter variants, `x` is passed as the
            tick value and `pos` is passed as the tick index.
        base : number
            Use log formatter (with scientific notation) having this value as the base.
            Set to `None` to override the default formatter with a log transform.
        unit : str or (str, str) tuple
            Use  SI prefixes with these units (e.g., with `unit="g"`, a tick value
            of 5000 will appear as `5 kg`). When a tuple, the first element gives the
            separator between the number and unit.

        Returns
        -------
        scale
            Copy of self with new label configuration.

        """

```

Please complete the target function and do not output anything else. Do not attach any docstrings.
\end{minted}


\section{Cross-file contexts examples}\label{app: context_examples}

In this section, we present an example from the seaborn repository (repository ID 0 in the REPOCOD benchmark) to illustrate the cross-file contexts retrieved by different methods: BM25, RepoMap, SpecAgent, and the Oracle Agent. The target function in this case is \texttt{Continuous.label}, whose ground-truth implementation is shown below:
\begin{minted}[fontsize=\scriptsize, breaklines, frame=single,  bgcolor=gray!5, breakanywhere]{python}
def label(
    self,
    formatter: Formatter | None = None, *,
    like: str | Callable | None = None,
    base: int | None | Default = default,
    unit: str | None = None,
) -> Continuous:
    """
    Configure the appearance of tick labels for the scale's axis or legend.
    
    ..."""
    # Input checks
    if formatter is not None and not isinstance(formatter, Formatter):
        raise TypeError(
            f"Label formatter must be an instance of {Formatter!r}, "
            f"not {type(formatter)!r}"
        )
    if like is not None and not (isinstance(like, str) or callable(like)):
        msg = f"`like` must be a string or callable, not {type(like).__name__}."
        raise TypeError(msg)
    
    new = copy(self)
    new._label_params = {
        "formatter": formatter,
        "like": like,
        "base": base,
        "unit": unit,
    }
    return new
\end{minted}

\subsection{BM25 contexts}

The cross-file contexts retrieved by BM25 are presented below, consisting of code chunks along with their respective relative file paths. There are a total of 12 context blocks, but only the initial two are displayed, and the remaining blocks are omitted due to their substantial length.
\begin{minted}[fontsize=\scriptsize, breaklines, frame=single,  bgcolor=gray!5, breakanywhere]{python}
# Code from: seaborn/categorical.py
    width=dedent("""\
    width : float
        Width allotted to each element on the orient axis. When `native_scale=True`,
        it is relative to the minimum distance between two values in the native scale.\
    """),
    dodge=dedent("""\
    dodge : "auto" or bool
        When hue mapping is used, whether elements should be narrowed and shifted along
        the orient axis to eliminate overlap. If `"auto"`, set to `True` when the
        orient variable is crossed with the categorical variable or `False` otherwise.
        .. versionchanged:: 0.13.0
            Added `"auto"` mode as a new default.\
    """),
    linewidth=dedent("""\
    linewidth : float
        Width of the lines that frame the plot elements.\
    """),
    linecolor=dedent("""\
    linecolor : color
        Color to use for line elements, when `fill` is True.
        .. versionadded:: v0.13.0\
    """),
    log_scale=dedent("""\
    log_scale : bool or number, or pair of bools or numbers
        Set axis scale(s) to log. A single value sets the data axis for any numeric
        axes in the plot. A pair of values sets each axis independently.
        Numeric values are interpreted as the desired base (default 10).
        When `None` or `False`, seaborn defers to the existing Axes scale.
        .. versionadded:: v0.13.0\
    """),
    native_scale=dedent("""\
    native_scale : bool
        When True, numeric or datetime values on the categorical axis will maintain
        their original scaling rather than being converted to fixed indices.
        .. versionadded:: v0.13.0\
    """),
    formatter=dedent("""\
    formatter : callable
        Function for converting categorical data into strings. Affects both grouping
        and tick labels.
        .. versionadded:: v0.13.0\
    """),
    legend=dedent("""\
    legend : "auto", "brief", "full", or False
        How to draw the legend. If "brief", numeric `hue` and `size`
        variables will be represented with a sample of evenly spaced values.
        If "full", every group will get an entry in the legend. If "auto",
        choose between brief or full representation based on number of levels.
        If `False`, no legend data is added and no legend is drawn.
        .. versionadded:: v0.13.0\
    """),

# Code from: seaborn/_core/plot.py
"variables": variables,
            "structure": structure,
            "wrap": wrap,
        }
        new = self._clone()
        new._facet_spec.update(spec)
        return new
    # TODO def twin()?
    def scale(self, **scales: Scale) -> Plot:
        """
        Specify mappings from data units to visual properties.
        Keywords correspond to variables defined in the plot, including coordinate
        variables (`x`, `y`) and semantic variables (`color`, `pointsize`, etc.).
        A number of "magic" arguments are accepted, including:
            - The name of a transform (e.g., `"log"`, `"sqrt"`)
            - The name of a palette (e.g., `"viridis"`, `"muted"`)
            - A tuple of values, defining the output range (e.g. `(1, 5)`)
            - A dict, implying a :class:`Nominal` scale (e.g. `{"a": .2, "b": .5}`)
            - A list of values, implying a :class:`Nominal` scale (e.g. `["b", "r"]`)
        For more explicit control, pass a scale spec object such as :class:`Continuous`
        or :class:`Nominal`. Or pass `None` to use an "identity" scale, which treats
        data values as literally encoding visual properties.
        Examples
        --------
        .. include:: ../docstrings/objects.Plot.scale.rst
        """
        new = self._clone()
        new._scales.update(scales)
        return new
    def share(self, **shares: bool | str) -> Plot:
        """
        Control sharing of axis limits and ticks across subplots.
        Keywords correspond to variables defined in the plot, and values can be
        boolean (to share across all subplots), or one of "row" or "col" (to share
        more selectively across one dimension of a grid).
        Behavior for non-coordinate variables is currently undefined.
        Examples
        --------
        .. include:: ../docstrings/objects.Plot.share.rst
        """
        new = self._clone()
        new._shares.update(shares)
        return new
    def limit(self, **limits: tuple[Any, Any]) -> Plot:
        """
        Control the range of visible data.
        Keywords correspond to variables defined in the plot, and values are a
        `(min, max)` tuple (where either can be `None` to leave unset).
        Limits apply only to the axis; data outside the visible range are
        still used for any stat transforms and added to the plot.
        """
\end{minted}

\subsection{RepoMap context}

The RepoMap context is presented below. Due to its considerable length, we only present a limited portion of the RepoMap context.
\begin{minted}[fontsize=\scriptsize, breaklines, frame=single,  bgcolor=gray!5, breakanywhere]{python}
# We provide you with structures of files that are imported by this target file, which only include their structure names such as global variable, class and function names, and their code implementations are omitted.
# These structures can help you understand the overall structure of imported files, and the relationships between the target file and its dependencies.
# For each imported file, we provide you with its file name, followed by its structure.

# seaborn/_core/typing.py
ColumnName
Vector
VariableSpec
VariableSpecList
DataSource
OrderSpec
NormSpec
PaletteSpec
DiscreteValueSpec
ContinuousValueSpec
class Default:
    def __repr__(self):
class Deprecated:
    def __repr__(self):
default
deprecated

# seaborn/_core/rules.py
class VarType(UserString):
    allowed
    def __init__(self, data):
    def __eq__(self, other):
def variable_type(
    vector: Series,
    boolean_type: Literal["numeric", "categorical", "boolean"] = "numeric",
    strict_boolean: bool = False,
) -> VarType:
    def all_numeric(x):
    def all_datetime(x):
def categorical_order(vector: Series, order: list | None = None) -> list:

# seaborn/_core/plot.py
default
class Layer(TypedDict, total=False):
    mark: Mark
    stat: Stat | None
    move: Move | list[Move] | None
    data: PlotData
    source: DataSource
    vars: dict[str, VariableSpec]
    orient: str
    legend: bool
    label: str | None
class FacetSpec(TypedDict, total=False):
    variables: dict[str, VariableSpec]
    structure: dict[str, list[str]]
    wrap: int | None
class PairSpec(TypedDict, total=False):
    variables: dict[str, VariableSpec]
    structure: dict[str, list[str]]
    cross: bool
    wrap: int | None
def theme_context(params: dict[str, Any]) -> Generator:
def build_plot_signature(cls):
...

# seaborn/_core/properties.py
RGBTuple
RGBATuple
ColorSpec
DashPattern
DashPatternWithOffset
MarkerPattern
Mapping
class Property:
    legend
    normed
    def __init__(self, variable: str | None = None):
    def default_scale(self, data: Series) -> Scale:
    def infer_scale(self, arg: Any, data: Series) -> Scale:
    def get_mapping(self, scale: Scale, data: Series) -> Mapping:
        def identity(x):
    def standardize(self, val: Any) -> Any:
    def _check_dict_entries(self, levels: list, values: dict) -> None:
    def _check_list_length(self, levels: list, values: list) -> list:
class Coordinate(Property):
    legend
    normed
...
\end{minted}

\subsection{Retriever agent contexts}

The cross-file contexts retrieved by the retriever agent are shown below. There are a total of 12 context blocks, but only the initial three are displayed, and the remaining blocks are omitted due to their substantial length.
\begin{minted}[fontsize=\scriptsize, breaklines, frame=single, bgcolor=gray!5, breakanywhere]{markdown}
An intelligent agent analyzed this repository and extracted the following relevant context for completing the target function.
The agent explored the codebase and identified these contexts as potentially useful based on patterns, dependencies, and contextual relevance.

Context 1: Looking at the base Scale class label method - this is the parent method that our implementation should be similar to
Type: code_snippet
The `label()` method in the Scale base class is a compatibility method currently implemented as a no-op:

```python
def label(self):
    # This is a no-op method for compatibility
    new = copy(self)
    new._label_params = {}
    return new
```

This suggests that our implementation should follow this pattern of returning a copy of the scale with updated parameters.

Context 2: Understanding how parameters are passed from label() method to _get_formatter()
Type: code_snippet
The Scale class's _get_scale method shows how label parameters are used:

```python
def _get_scale(self, name: str, forward: Callable, inverse: Callable):
    major_locator, minor_locator = self._get_locators(**self._tick_params)
    major_formatter = self._get_formatter(major_locator, **self._label_params)
    
    class InternalScale(mpl.scale.FuncScale):
        def set_default_locators_and_formatters(self, axis):
            axis.set_major_locator(major_locator)
            if minor_locator is not None:
                axis.set_minor_locator(minor_locator)
            axis.set_major_formatter(major_formatter)
    
    return InternalScale(name, (forward, inverse))
```

This shows that `self._label_params` are passed as kwargs to the _get_formatter method.

Context 3: Examining Continuous.tick() method signature and implementation
Type: code_snippet
The `Continuous.tick()` method is highly relevant as it shows how to implement an interface for configuring ticks:

```python
def tick(
    self,
    locator: Locator | None = None, *,
    at: Sequence[float] | None = None,
    upto: int | None = None,
    count: int | None = None,
    every: float | None = None,
    between: tuple[float, float] | None = None,
    minor: int | None = None,
) -> Continuous:
    """
    Configure the selection of ticks for the scale's axis or legend.

    Parameters
    ----------
    locator : :class:`matplotlib.ticker.Locator` subclass
        Pre-configured matplotlib locator; other parameters will not be used.
    at : sequence of floats
        Place ticks at these specific locations (in data units).
    upto : int
        Choose "nice" locations for ticks, but do not exceed this number.
    count : int
        Choose exactly this number of ticks, bounded by `between` or axis limits.
    every : float
        Choose locations at this interval of separation (in data units).
    between : pair of floats
        Bound upper / lower ticks when using `every` or `count`.
    minor : int
        Number of unlabeled ticks to draw between labeled "major" ticks.

    Returns
    -------
    scale
        Copy of self with new tick configuration.

    """
    # Input checks
    if locator is not None and not isinstance(locator, Locator):
        raise TypeError(
            f"Tick locator must be an instance of {Locator!r}, "
            f"not {type(locator)!r}."
        )
    log_base, symlog_thresh = self._parse_for_log_params(self.trans)
    if log_base or symlog_thresh:
        if count is not None and between is None:
            raise RuntimeError("`count` requires `between` with log transform.")
        if every is not None:
            raise RuntimeError("`every` not supported with log transform.")

    new = copy(self)
    new._tick_params = {
        "locator": locator,
        "at": at,
        "upto": upto,
        "count": count,
        "every": every,
        "between": between,
        "minor": minor,
    }
    return new
```

This pattern follows the same approach - validate parameters, create a copy of self, update parameter dictionary, and return new instance.
\end{minted}

\subsection{Forecaster agent}

The cross-file contexts retrieved by the forecaster agent are shown below. The prediction of the forecaster agent is incorrect for this sample; the target function is \texttt{label}, but the agent predicts \texttt{\_default\_spacer}. We only show the initial three context blocks, and the remaining blocks are omitted due to their substantial length.
\begin{minted}[fontsize=\scriptsize, breaklines, frame=single,  bgcolor=gray!5, breakanywhere]{markdown}
An intelligent agent analyzed this repository and extracted the following relevant context for completing the target function.
The agent explored the codebase and identified these contexts as potentially useful based on patterns, dependencies, and contextual relevance.

Context 1: This implementation provides a simple function that returns a default spacing value of 1 for any input series. This is consistent with the Scale._spacing method's fallback behavior when there's no variance in the data.
Source: predicted_function
Type: predicted_implementation
```python
def _default_spacer(x: Series) -> float:
    """
    Return a default spacing value of 1 for any input.
    
    This function is used as a default spacing function for scales where 
    specific spacing behavior is not needed.
    
    Parameters
    ----------
    x : Series
        Input data (not used in this implementation)
    
    Returns
    -------
    float
        Default spacing value (1.0)
    """
    return 1.0
```

Context 2: This implementation calculates spacing based on the range of the data divided by a standard number of bins (10). This approach provides spacing proportional to the data range while ensuring a reasonable number of divisions.
Source: predicted_function
Type: predicted_implementation
```python
def _default_spacer(x: Series) -> float:
    """
    Calculate a default spacing value based on the data range.
    
    Returns the range of the data divided by 10 as a reasonable default spacing.
    If the data contains fewer than 2 unique values or is all NaN, returns 1.0.
    
    Parameters
    ----------
    x : Series
        Input data series
    
    Returns
    -------
    float
        Calculated spacing value
    """
    clean_x = x.dropna()
    if len(clean_x) < 2:
        return 1.0
    
    x_min, x_max = clean_x.min(), clean_x.max()
    if x_min == x_max:
        return 1.0
        
    return (x_max - x_min) / 10
```

Context 3: This implementation is similar to how the spacing is handled in the existing Continuous scale class, using the minimum difference between sorted values to determine natural spacing.
Source: predicted_function
Type: predicted_implementation
```python
def _default_spacer(x: Series) -> float:
    """
    Calculate spacing based on the minimum difference between adjacent values.
    
    Finds the minimum difference between adjacent values in the sorted data,
    which is useful for determining appropriate bar widths or point spacing.
    
    Parameters
    ----------
    x : Series
        Input data series
    
    Returns
    -------
    float
        Minimum difference between adjacent values or 1.0 if undetermined
    """
    clean_x = x.dropna().unique()
    if len(clean_x) < 2:
        return 1.0
    
    sorted_x = np.sort(clean_x)
    min_diff = np.min(np.diff(sorted_x))
    
    # Guard against zero spacing
    return max(min_diff, 1e-8) if min_diff > 0 else 1.0
```
\end{minted}

\subsection{SpecAgent contexts}

The cross-file contexts retrieved by the SpecAgent are shown below. Interestingly, the SpecAgent's prediction for the target function is incorrect in this example; it predicts \texttt{\_make\_log\_transforms}, but the real target function is actually \texttt{Continuous.scale}. We only show the initial three context blocks, and the remaining blocks are omitted due to their substantial length.
\begin{minted}[fontsize=\scriptsize, breaklines, frame=single,  bgcolor=gray!5, breakanywhere]{markdown}
An intelligent agent analyzed this repository and extracted the following relevant context for completing the target function.
The agent explored the codebase and identified these contexts as potentially useful based on patterns, dependencies, and contextual relevance.

Context 1: The most likely missing function is `_make_log_transforms`. This function creates forward and inverse transform functions for logarithmic scaling. Looking at the imports and existing transforms, here's the implementation:
Type: code_snippet
```python
def _make_log_transforms(base: float | None = None) -> TransFuncs:

    fs: TransFuncs
    if base is None:
        fs = np.log, np.exp
    elif base == 2:
        fs = np.log2, partial(np.power, 2)
    elif base == 10:
        fs = np.log10, partial(np.power, 10)
    else:
        def forward(x):
            return np.log(x) / np.log(base)
        fs = forward, partial(np.power, base)

    def log(x: ArrayLike) -> ArrayLike:
        with np.errstate(invalid="ignore", divide="ignore"):
            return fs[0](x)

    def exp(x: ArrayLike) -> ArrayLike:
        with np.errstate(invalid="ignore", divide="ignore"):
            return fs[1](x)

    return log, exp
```

This function is used extensively throughout the codebase for logarithmic transformations with different bases.

Context 2: Function usage in `ContinuousBase._get_transform` method in `seaborn/_core/scales.py`:
Source: seaborn/_core/scales.py
Type: code_snippet
The `_make_log_transforms` function is used in multiple places in the `_get_transform` method. Here are the relevant parts:

```python
def _get_transform(self):
    arg = self.trans
    def get_param(method, default):
        if arg == method:
            return default
        return float(arg[len(method):])

    if arg is None:
        return _make_identity_transforms()
    elif isinstance(arg, tuple):
        return arg
    elif isinstance(arg, str):
        if arg == "ln":
            return _make_log_transforms()
        elif arg == "logit":
            base = get_param("logit", 10)
            return _make_logit_transforms(base)
        elif arg.startswith("log"):
            base = get_param("log", 10)
            return _make_log_transforms(base)
        # Other transform options...
```

This shows that `_make_log_transforms` is called when the transformation type is "ln" or when it starts with "log". The function supports both natural logarithm and logarithms with specific bases.

Context 3: Usage in other transformation functions in `seaborn/_core/scales.py`:
Source: seaborn/_core/scales.py
Type: code_snippet
The `_make_log_transforms` function is also used within other transform functions:

```python
def _make_logit_transforms(base: float | None = None) -> TransFuncs:
    log, exp = _make_log_transforms(base)
    
    def logit(x):
        with np.errstate(invalid="ignore", divide="ignore"):
            return log(x) - log(1 - x)
    
    def expit(x):
        with np.errstate(invalid="ignore", divide="ignore"):
            return exp(x) / (1 + exp(x))
    
    return logit, expit

def _make_symlog_transforms(c: float = 1, base: float = 10) -> TransFuncs:
    # From https://iopscience.iop.org/article/10.1088/0957-0233/24/2/027001
    log, exp = _make_log_transforms(base)
    
    def symlog(x):
        with np.errstate(invalid="ignore", divide="ignore"):
            return np.sign(x) * log(1 + np.abs(np.divide(x, c)))
    
    def symexp(x):
        with np.errstate(invalid="ignore", divide="ignore"):
            return np.sign(x) * c * (exp(np.abs(x)) - 1)
    
    return symlog, symexp
```

This shows that `_make_log_transforms` serves as a building block for more complex transformations.
\end{minted}

\subsection{Oracle agent contexts}

The contexts retrieved by the oracle agent are shown below. Notably, while the oracle agent has access to the ground-truth function completion, it does not provide a direct solution in the contexts. There are a total of 12 context blocks, but only the initial three are displayed, and the remaining blocks are omitted due to their substantial length.
\begin{minted}[fontsize=\scriptsize, breaklines, frame=single, bgcolor=gray!5, breakanywhere]{markdown}
An intelligent agent analyzed this repository and extracted the following relevant context for completing the target function.
The agent explored the codebase and identified these contexts as potentially useful based on patterns, dependencies, and contextual relevance.

Context 1: Implementation of _get_formatter method for Continuous scales that handles formatter logic
Source: seaborn/_core/scales.py
Type: code_snippet
From the file seaborn/_core/scales.py, the Continuous class has a _get_formatter method that shows how formatters are created and configured based on parameters:

```python
def _get_formatter(self, locator, formatter, like, base, unit):
    log_base, symlog_thresh = self._parse_for_log_params(self.trans)
    if base is default:
        if symlog_thresh:
            log_base = 10
        base = log_base

    if formatter is not None:
        return formatter

    if like is not None:
        if isinstance(like, str):
            if "{x" in like or "{pos" in like:
                fmt = like
            else:
                fmt = f"{{x:{like}}}"
            formatter = StrMethodFormatter(fmt)
        else:
            formatter = FuncFormatter(like)

    elif base is not None:
        # We could add other log options if necessary
        formatter = LogFormatterSciNotation(base)

    elif unit is not None:
        if isinstance(unit, tuple):
            sep, unit = unit
        elif not unit:
            sep = ""
        else:
            sep = " "
        formatter = EngFormatter(unit, sep=sep)

    else:
        formatter = ScalarFormatter()

    return formatter
```

Context 2: Temporal class label method showing similar pattern
Source: seaborn/_core/scales.py
Type: code_snippet
The Temporal class's label method in seaborn/_core/scales.py shows a similar pattern for configuring tick label formatters:

```python
def label(
    self,
    formatter: Formatter | None = None, *,
    concise: bool = False,
) -> Temporal:
    """
    Configure the appearance of tick labels for the scale's axis or legend.

    .. note::
        This API is under construction and will be enhanced over time.

    Parameters
    ----------
    formatter : :class:`matplotlib.ticker.Formatter` subclass
        Pre-configured formatter to use; other parameters will be ignored.
    concise : bool
        If True, use :class:`matplotlib.dates.ConciseDateFormatter` to make
        the tick labels as compact as possible.

    Returns
    -------
    scale
        Copy of self with new label configuration.

    """
    new = copy(self)
    new._label_params = {"formatter": formatter, "concise": concise}
    return new
```

Context 3: Boolean class label method implementation
Source: seaborn/_core/scales.py
Type: code_snippet
From seaborn/_core/scales.py, the Boolean class shows a simpler implementation of the label method:

```python
def label(self, formatter: Formatter | None = None):
    new = copy(self)
    new._label_params = {"formatter": formatter}
    return new
```
\end{minted}


\section{Retrieval methods cross-comparisons}

In this section, we compare traditional retrieval methods commonly used in production systems (BM25 and RepoMap) with the oracle agent described in \Cref{sec: method_oracle}. Our analysis focuses on cases where one method leads to a correct function completion (i.e., passes unit tests) while another fails. In particular, we investigate whether these discrepancies stem from missing or insufficient cross-file context---cases we term \emph{context decisive}.

To make this determination, we use an LLM judge to assess why one method succeeds while the other does not, and whether the difference can be attributed to context decisiveness. From these results, we derive insights into why BM25 and RepoMap underperform compared to the oracle agent, which in turn inform the design of our indexing-time agents to mitigate these limitations. We also provide representative examples for each insight. For clarity, the ``oracle agent'' referenced in this section denotes an earlier version of the agent, which is less performant than the one reported in the main text.

\subsection{Evaluation setup}

For each REPOCOD sample in which one cross-file retrieval method succeeds while the other fails, we ask an LLM judge to analyze the cause of this discrepancy. Specifically, the judge assesses (i) the likely reason for the outcome (e.g., context quality, missing imports, wrong API usage, or logic errors), (ii) whether the difference is \emph{context decisive}, and (iii) whether the retrieved contexts were helpful for function completion. We employ Claude 3.7 Sonnet~\citep{claude-3-7} as the judge, and provide it with a structured prompt. The exact prompt format is shown below.
\begin{minted}[fontsize=\scriptsize, breaklines, frame=single, bgcolor=gray!5, breakanywhere]{markdown}
You are an expert software engineer analyzing why two different code completion methods produced different results on the same task.

**Context:**
- Repository: {repo_name}
- Function: {function_name}
- One method succeeded (passed unit tests) while the other failed
- Success method: {success_method.upper()}

**Task Information:**

**1. Original prompt given to both models:**
```
{prompt}
```

**2. Target file content (for context):**
```python
{target_file_content[:2000]}{"..." if len(target_file_content) > 2000 else ""}
```

**3. Ground truth (correct implementation):**
```python
{ground_truth}
```

**4. BM25 Method Output:**
```python
{bm25_output}
```

**5. Oracle Agent Method Output:**
```python
{oracle_output}
```

**6. Cross-file contexts used by each method:**

{bm25_context_str}

{oracle_context_str}

**Analysis Task:**
Analyze why the {success_method.upper()} method succeeded while the other failed. Focus on:

1. **Context Quality**: How did the different cross-file contexts influence the results?
2. **Missing Information**: What critical information was missing in the failing method's context?
3. **Context Relevance**: Which context was more relevant to solving the task?
4. **Specific Differences**: What specific differences in the outputs led to success/failure?

**Response Format:**
Provide a JSON response with:
{{
    "success_method": "{success_method}",
    "primary_difference_reason": "context_quality|missing_imports|wrong_api_usage|logic_error|other",
    "context_was_decisive": true/false,
    "bm25_context_assessment": "helpful|partially_helpful|unhelpful|missing",
    "oracle_context_assessment": "helpful|partially_helpful|unhelpful|missing", 
    "detailed_analysis": "Detailed explanation of why one method succeeded",
    "critical_missing_context": "What context was missing in the failed method",
    "context_quality_comparison": "Compare the quality and relevance of both contexts",
    "confidence": "high|medium|low"
}}

Only respond with the JSON, no additional text, no markdown code blocks.
\end{minted}

\subsection{BM25 versus oracle agent}

\subsubsection{Summary}

The LLM judge analysis reveals clear differences between BM25 and the oracle agent:
\begin{itemize}
    \item The oracle agent achieves \textbf{98.8\% context decisiveness}, compared to BM25's \textbf{25.5\%}, indicating that oracle's successes are far more often driven by relevant cross-file context.
    \item Among all samples analyzed, the oracle agent provides \textbf{helpful context in 100\% of cases}, whereas BM25 does so in only \textbf{21.6\% of cases}.
\end{itemize}

As shown in \Cref{fig: bm25_vs_oracle_breakdown}, the oracle agent's advantage primarily stems from its ability to retrieve high-quality cross-file context, which directly enables correct function completions. In contrast, BM25's rare successes tend to arise from logic-related issues where context quality plays little to no role. These results underscore that context retrieval is the decisive factor behind the oracle agent's superior performance.

In the following subsection, we present detailed case studies to better understand the common failure modes of BM25 and why it often fails to surface the most relevant context.

\subsubsection{Insights}\label{app: bm25_vs_oracle_insights}

In this subsection, we present a set of detailed insights into why BM25 often fails to retrieve the most helpful contexts, while the oracle agent avoids these pitfalls. For each insight, we provide concrete examples drawn from our experiments. These examples demonstrate how the oracle agent's access to richer and semantically aligned information enables more accurate code completion, while BM25's reliance on sparse lexical similarity leads to systematic blind spots. Together, these insights highlight the advantages of using an intelligent retrieval agent over traditional methods based purely on token similarity.

A breakdown of the reasons behind each method's successful completions is shown in \Cref{fig: bm25_vs_oracle_breakdown}.
\begin{figure}[htbp]
    \centering
    \begin{subfigure}[b]{\columnwidth}
        \centering
        \begin{adjustbox}{max width=\textwidth}
            \begin{tikzpicture}
                \pie[sum=auto, text=legend]{41/context quality, 22/logic error, 7/missing information, 7/wrong API usage, 7/missing imports}
            \end{tikzpicture}
        \end{adjustbox}
        \caption{Reasons for oracle agent successes. Context quality is the dominant factor.}
    \end{subfigure}
    \hfill
    \begin{subfigure}[b]{\columnwidth}
        \centering
        \begin{adjustbox}{max width=\textwidth}
            \begin{tikzpicture}
                \pie[sum=auto, text=legend]{31/logic error, 14/wrong API usage, 4/context quality, 2/missing information}
            \end{tikzpicture}
        \end{adjustbox}
        \caption{Reasons for BM25 successes. Logic errors are most common.}
    \end{subfigure}
    \caption{Breakdown of success factors for BM25 and the oracle agent. Oracle successes are usually context-driven, while BM25's are often due to issues unrelated to context quality.}
    \label{fig: bm25_vs_oracle_breakdown}
\end{figure}

\begin{enumerate}
    \item \textbf{Insight}: The ground-truth implementation may invoke other functions in the repository that are semantically important but lexically unrelated. Since BM25 relies on surface similarity in terms of docstrings and function signatures, it frequently fails to retrieve these functions. The oracle agent, by contrast, has access to the target function body and therefore prioritizes the contexts of functions that are directly invoked.

    \textbf{Example}: In repository ID 67 from the seaborn repository, the target function is \texttt{tick} in the \texttt{Continuous} class. The function internally calls \texttt{\_parse\_for\_log\_params}. The oracle agent retrieves the definition and usages of this helper function, allowing the completion model to include the correct call in its generated output:
    \begin{minted}[fontsize=\scriptsize, breaklines, frame=single, bgcolor=gray!5, escapeinside=~~]{python}
def tick(
    self,
    locator: Locator | None = None, *,
    at: Sequence[float] | None = None,
    upto: int | None = None,
    count: int | None = None,
    every: float | None = None,
    between: tuple[float, float] | None = None,
    minor: int | None = None,
) -> Continuous:
    if locator is not None and not isinstance(locator, Locator):
        err = (
            f"Tick locator must be an instance of {Locator!r}, "
            f"not {type(locator)!r}."
        )
        raise TypeError(err)

    log_base, symlog_thresh = self.~\bfseries{~_parse_for_log_params~}~(self.trans)

    if log_base or symlog_thresh:
        if count is not None and between is None:
            raise RuntimeError("`count` requires `between` with log transform.")
        if every is not None:
            raise RuntimeError("`every` not supported with log transform.")

    new = copy(self)
    new._tick_params = {
        "locator": locator,
        "at": at,
        "upto": upto,
        "count": count,
        "every": every,
        "between": between,
        "minor": minor,
    }
    return new
    \end{minted}

    Contexts related to this helper function are included in the following two blocks in the oracle agent's retrieved contexts:
    \begin{minted}[fontsize=\scriptsize, breaklines, frame=single, bgcolor=gray!5, breakanywhere]{markdown}
**Explanation**: Log parameter detection in `Continuous._parse_for_log_params` method

From `seaborn/_core/scales.py`, the `Continuous._get_locators` method shows the logic for selecting different locators based on parameters:

```python
def _get_locators(self, locator, at, upto, count, every, between, minor):
    log_base, symlog_thresh = self._parse_for_log_params(self.trans)

    if locator is not None:
        major_locator = locator

    elif upto is not None:
        if log_base:
            major_locator = LogLocator(base=log_base, numticks=upto)
        else:
            major_locator = MaxNLocator(upto, steps=[1, 1.5, 2, 2.5, 3, 5, 10])

    elif count is not None:
        if between is None:
            # This is rarely useful (unless you are setting limits)
            major_locator = LinearLocator(count)
        else:
            if log_base or symlog_thresh:
                forward, inverse = self._get_transform()
                lo, hi = forward(between)
                ticks = inverse(np.linspace(lo, hi, num=count))
            else:
                ticks = np.linspace(*between, num=count)
            major_locator = FixedLocator(ticks)

    elif every is not None:
        if between is None:
            major_locator = MultipleLocator(every)
        else:
            lo, hi = between
            ticks = np.arange(lo, hi + every, every)
            major_locator = FixedLocator(ticks)

    elif at is not None:
        major_locator = FixedLocator(at)

    else:
        if log_base:
            major_locator = LogLocator(log_base)
        elif symlog_thresh:
            major_locator = SymmetricalLogLocator(linthresh=symlog_thresh, base=10)
        else:
            major_locator = AutoLocator()

    if minor is None:
        minor_locator = LogLocator(log_base, subs=None) if log_base else None
    else:
        if log_base:
            subs = np.linspace(0, log_base, minor + 2)[1:-1]
            minor_locator = LogLocator(log_base, subs=subs)
        else:
            minor_locator = AutoMinorLocator(minor + 1)

    return major_locator, minor_locator
```

**Explanation**: Log parameter detection in `Continuous._parse_for_log_params` method

From `seaborn/_core/scales.py`, the `_parse_for_log_params` method shows how to detect log and symlog parameters:

```python
def _parse_for_log_params(
    self, trans: str | TransFuncs | None
) -> tuple[float | None, float | None]:

    log_base = symlog_thresh = None
    if isinstance(trans, str):
        m = re.match(r"^log(\d*)", trans)
        if m is not None:
            log_base = float(m[1] or 10)
        m = re.match(r"symlog(\d*)", trans)
        if m is not None:
            symlog_thresh = float(m[1] or 1)
    return log_base, symlog_thresh
```
    \end{minted}

    With this contextual support, the generated code passes the unit tests. By contrast, BM25 retrieves contexts unrelated to \texttt{\_parse\_for\_log\_params}, as there is no lexical similarity between its name or docstring and the target function. Consequently, the model fails to call this helper function, and the resulting completion is incorrect. This example illustrates a structural weakness of BM25: it cannot capture dependency chains that are critical for correctness.
    \item \label{it: bm25_vs_oracle_insights_conventions}\textbf{Insight}: BM25 often overlooks repository-wide conventions that shape function behavior. Such conventions include default fallback values, strategies for handling missing inputs, and typical patterns of error handling. Since these conventions are rarely reflected in surface-level lexical overlap, BM25 is blind to them. The oracle agent, however, surfaces conventions by retrieving semantically related examples and type information.

    \textbf{Example}: In repository ID 15 from more-itertools, the target function is \texttt{divide}, which should return a list of iterators. The target's signature provides no type information. BM25 misses the type hints in \texttt{more.pyi} specifying the return type as \texttt{list[Iterator[\_T]]}, and it fails to bring in related functions such as \texttt{distribute()} that demonstrate the correct use of iterators. The oracle agent, by contrast, retrieves the type stubs, similar functions such as \texttt{chunked\_even}, and examples of using \texttt{divmod} for element partitioning:
    \begin{minted}[fontsize=\scriptsize, breaklines, frame=single, bgcolor=gray!5, breakanywhere]{markdown}
**Explanation**: The `chunked_even` function shows dividing an iterable into evenly sized chunks

From `more_itertools/more_itertools/more.py`:
```python
def chunked_even(iterable, n):
    """Break *iterable* into lists of approximately length *n*.
    Items are distributed such the lengths of the lists differ by at most
    1 item.

    >>> iterable = [1, 2, 3, 4, 5, 6, 7]
    >>> n = 3
    >>> list(chunked_even(iterable, n))  # List lengths: 3, 2, 2
    [[1, 2, 3], [4, 5], [6, 7]]
    >>> list(chunked(iterable, n))  # List lengths: 3, 3, 1
    [[1, 2, 3], [4, 5, 6], [7]]

    """
    iterable = iter(iterable)

    # Initialize a buffer to process the chunks while keeping
    # some back to fill any underfilled chunks
    min_buffer = (n - 1) * (n - 2)
    buffer = list(islice(iterable, min_buffer))

    # Append items until we have a completed chunk
    for _ in islice(map(buffer.append, iterable), n, None, n):
        yield buffer[:n]
        del buffer[:n]

    # Check if any chunks need addition processing
    if not buffer:
        return
    length = len(buffer)

    # Chunks are either size `full_size <= n` or `partial_size = full_size - 1`
    q, r = divmod(length, n)
    num_lists = q + (1 if r > 0 else 0)
    q, r = divmod(length, num_lists)
    full_size = q + (1 if r > 0 else 0)
    partial_size = full_size - 1
    num_full = length - partial_size * num_lists

    # Yield chunks of full size
    partial_start_idx = num_full * full_size
    if full_size > 0:
        for i in range(0, partial_start_idx, full_size):
            yield buffer[i : i + full_size]

    # Yield chunks of partial size
    if partial_size > 0:
        for i in range(partial_start_idx, length, partial_size):
            yield buffer[i : i + partial_size]
```

**Explanation**: The function type signature for `divide` from `more.pyi` (based on found imports)

From `more_itertools/more_itertools/more.pyi`:
```python
def divide(n: int, iterable: Iterable[_T]) -> list[Iterator[_T]]: ...
```
    \end{minted}

    Equipped with these contexts, the model produces a correct implementation that returns iterators rather than lists. Without them, BM25's contexts bias the model toward incorrect type usage. This example underscores the role of semantic retrieval in preserving repository-specific coding conventions.
    \item \textbf{Insight}: Many repositories exhibit ``multi-hop context'' dependencies, where the correct implementation requires knowledge not only of the immediate class APIs but also of the APIs of member objects within that class. BM25 generally fails to capture these multi-hop relationships, as they involve broader structural reasoning across classes. The oracle agent, by design, retrieves relevant cross-references and usage patterns.

    \textbf{Example}: In repository ID 26 from seaborn, the target function is \texttt{refline} in the \texttt{JointGrid} class. The code completion model with access to the oracle agent's contexts generates the following correct function implementation:
    \begin{minted}[fontsize=\scriptsize, breaklines, frame=single, bgcolor=gray!5, escapeinside=||, breakanywhere]{python}
def refline(
    self, *, x=None, y=None, joint=True, marginal=True,
    color='.5', linestyle='--', **line_kws
):
    """Add a reference line(s) to joint and/or marginal axes.

    Parameters
    ----------
    x, y : numeric
        Value(s) to draw the line(s) at.
    joint, marginal : bools
        Whether to add the reference line(s) to the joint/marginal axes.
    color : :mod:`matplotlib color <matplotlib.colors>`
        Specifies the color of the reference line(s).
    linestyle : str
        Specifies the style of the reference line(s).
    line_kws : key, value mappings
        Other keyword arguments are passed to :meth:`matplotlib.axes.Axes.axvline`
        when ``x`` is not None and :meth:`matplotlib.axes.Axes.axhline` when ``y``
        is not None.

    Returns
    -------
    :class:`JointGrid` instance
        Returns ``self`` for easy method chaining.

    """
    line_kws['color'] = color
    line_kws['linestyle'] = linestyle

    if x is not None and joint:
        self.ax_joint.axvline(x=x, **line_kws)
    if y is not None and joint:
        self.ax_joint.axhline(y=y, **line_kws)
    if x is not None and marginal:
        self.|\bfseries{|ax_marg_x|}|.axvline(x=x, **line_kws)
    if y is not None and marginal:
        self.|\bfseries{|ax_marg_y|}|.axhline(y=y, **line_kws)

    return self
    \end{minted}

    Clearly, to generate a correct completion, the model must understand the roles of the \texttt{ax\_marg\_x} and \texttt{ax\_marg\_y} attributes, which are matplotlib axes objects. BM25 fails to retrieve detailed contexts about the structure of \texttt{JointGrid} or how its axes are used. The oracle agent, however, surfaces examples of how reference lines are added to axes, how the \texttt{JointGrid} object manages its subplots, and how marginal axes are manipulated:
    \begin{minted}[fontsize=\scriptsize, breaklines, frame=single, bgcolor=gray!5, breakanywhere]{markdown}
**Explanation**: `JointGrid` class property that gives access to the axes

From the code in the target file, we can see how `JointGrid`'s axes are structured:

```python
class JointGrid(_BaseGrid):
    # ...
    def __init__(
        self, data=None, *,
        x=None, y=None, hue=None,
        height=6, ratio=5, space=.2,
        palette=None, hue_order=None, hue_norm=None,
        dropna=False, xlim=None, ylim=None, marginal_ticks=False,
    ):
        # ...
        # Set up the subplot grid
        f = plt.figure(figsize=(height, height))
        gs = plt.GridSpec(ratio + 1, ratio + 1)

        ax_joint = f.add_subplot(gs[1:, :-1])
        ax_marg_x = f.add_subplot(gs[0, :-1], sharex=ax_joint)
        ax_marg_y = f.add_subplot(gs[1:, -1], sharey=ax_joint)

        self._figure = f
        self.ax_joint = ax_joint
        self.ax_marg_x = ax_marg_x
        self.ax_marg_y = ax_marg_y
```
    \end{minted}

    With this information, the completion model successfully generates the correct function. This example illustrates how semantic retrieval captures higher-order class dependencies that token-based similarity cannot.
    \item \textbf{Insight}: BM25 may provide partial coverage of API calls but often fails to retrieve contexts showing how different functions interact together in realistic workflows. The oracle agent excels at recovering these co-usage patterns, which are crucial for correctness.

    \textbf{Example}: In repository ID 28 from more-itertools, the target function is \texttt{locate}. Its implementation requires the combined use of \texttt{compress} and \texttt{count}:
    \begin{minted}[fontsize=\scriptsize, breaklines, frame=single, bgcolor=gray!5, escapeinside=||, breakanywhere]{python}
def locate(iterable, pred=bool, window_size=None):
    if window_size is None:
        try:
            return |\bfseries{|compress|}|(|\bfseries{|count|}|(), map(pred, iterable))
        except TypeError:
            pass
    if window_size < 1:
        raise ValueError('window_size must be at least 1')
    it = iter(iterable)
    windows = windowed(it, window_size)
    return |\bfseries{|compress|}|(|\bfseries{|count|}|(), map(lambda w: pred(*w), windows))
    \end{minted}

    BM25 retrieves fragments showing these functions individually but not in combination. As a result, the model lacks examples demonstrating their joint usage. By contrast, the oracle agent retrieves definitions, test cases, and helper functions showing how these two functions interact:
    \begin{minted}[fontsize=\scriptsize, breaklines, frame=single, bgcolor=gray!5, breakanywhere]{markdown}
**Explanation**: The pattern for combining `itertools.compress` with `count()` to generate indexes is seen in other functions

From `more_itertools/more_itertools/recipes.py`:
```python
def iter_index(iterable, value, start=0, stop=None):
    """Yield the index of each place in *iterable* that *value* occurs,
    beginning with index *start* and ending before index *stop*.


    >>> list(iter_index('AABCADEAF', 'A'))
    [0, 1, 4, 7]
    >>> list(iter_index('AABCADEAF', 'A', 1))  # start index is inclusive
    [1, 4, 7]
    >>> list(iter_index('AABCADEAF', 'A', 1, 7))  # stop index is not inclusive
    [1, 4]

    The behavior for non-scalar *values* matches the built-in Python types.

    >>> list(iter_index('ABCDABCD', 'AB'))
    [0, 4]
    >>> list(iter_index([0, 1, 2, 3, 0, 1, 2, 3], [0, 1]))
    []
    >>> list(iter_index([[0, 1], [2, 3], [0, 1], [2, 3]], [0, 1]))
    [0, 2]

    See :func:`locate` for a more general means of finding the indexes
    associated with particular values.

    """
    seq_index = getattr(iterable, 'index', None)
    if seq_index is None:
        # Slow path for general iterables
        it = islice(iterable, start, stop)
        for i, element in enumerate(it, start):
            if element is value or element == value:
                yield i
    else:
        # Fast path for sequences
        stop = len(iterable) if stop is None else stop
        i = start - 1
        try:
            while True:
                yield (i := seq_index(value, i + 1, stop))
        except ValueError:
            pass
```
    \end{minted}

    This richer context leads to a correct implementation, while BM25's incomplete view results in failure. The lesson here is that the retrieval method must capture not only components but also how they are orchestrated together.
    \item\label{it: bm25_vs_oracle_insights_semantically_similar} \textbf{Insight}: BM25 systematically struggles with retrieving semantically similar but lexically distinct functions. These functions may use different terminology or naming conventions, making lexical similarity an unreliable signal. The oracle agent is robust to this mismatch because it prioritizes semantic connections over token overlap.

    \textbf{Example}: In repository ID 6 from more-itertools, the target function is \texttt{classify\_unique}, which references \texttt{unique\_everseen} and \texttt{unique\_justseen} in its docstring:
    \begin{minted}[fontsize=\scriptsize, breaklines, frame=single, bgcolor=gray!5, escapeinside=||, breakanywhere]{python}
def classify_unique(iterable, key=None):
    seen_set = set()
    seen_list = []
    use_key = key is not None
    prev_key = None
    for element in iterable:
        k = key(element) if use_key else element
        try:
            if k not in seen_set:
                seen_set.add(k)
                is_everseen = True
            else:
                is_everseen = False
            if prev_key is None or k != prev_key:
                is_justseen = True
                prev_key = k
            else:
                is_justseen = False
            yield (element, is_justseen, is_everseen)
        except TypeError:
            if k not in seen_list:
                seen_list.append(k)
                is_everseen = True
            else:
                is_everseen = False
            if prev_key is None or k != prev_key:
                is_justseen = True
                prev_key = k
            else:
                is_justseen = False
            yield (element, |\bfseries{|is_justseen|}|, |\bfseries{|is_everseen|}|)
    \end{minted}

    BM25 fails to retrieve these related functions due to low lexical similarity, leaving the model without a complete picture. The oracle agent retrieves both, along with test cases that clarify the intended behavior:
    \begin{minted}[fontsize=\scriptsize, breaklines, frame=single, bgcolor=gray!5, breakanywhere]{markdown}
**Explanation**: The `unique_everseen` function from `recipes.py` would be highly relevant as the `classify_unique` function description suggests returning similar functionality.

From `more_itertools/recipes.py`:
```python
def unique_everseen(iterable, key=None):
    """
    List unique elements, preserving order. Remember all elements ever seen.

        >>> list(unique_everseen('AAAABBBCCDAABBB'))
        ['A', 'B', 'C', 'D']
        >>> list(unique_everseen('ABBCcAD', str.lower))
        ['a', 'b', 'c', 'd']

    Sequences with a mix of hashable and unhashable items can be used.
    The function will be slower (i.e., O(n^2)) for unhashable items.

    """
    seenset = set()
    seenlist = []
    use_key = key is not None
    for element in iterable:
        k = key(element) if use_key else element
        try:
            if k not in seenset:
                seenset.add(k)
                yield element
        except TypeError:
            if k not in seenlist:
                seenlist.append(k)
                yield element
```

**Explanation**: The `groupby` function from itertools is used in several places and could be relevant for detecting consecutive duplicates:

From `more_itertools/recipes.py`:
```python
def unique_justseen(iterable, key=None):
    """List unique elements, preserving order. Remember only the element just seen.

        >>> list(unique_justseen('AAAABBBCCDAABBB'))
        ['A', 'B', 'C', 'D', 'A', 'B']
        >>> list(unique_justseen('ABBCcAD', str.lower))
        ['a', 'b', 'c', 'a', 'd']

    """
    return map(next, map(itemgetter(1), groupby(iterable, key)))
```
    \end{minted}

    With this full context, the model generates a correct function. BM25's incomplete retrieval, by contrast, leads to missing logic and failed tests. This case exemplifies how token-based retrieval misfires when functional similarity is not aligned with naming similarity.
\end{enumerate}

\subsection{RepoMap versus oracle agent}

\subsubsection{Summary}

The LLM-judge evaluation highlights a stark contrast between RepoMap and the oracle agent:
\begin{itemize}
    \item The oracle agent achieves \textbf{97.8\% context decisiveness}, compared to RepoMap's \textbf{41.7\%}. This indicates that the oracle's completions are far more often enabled by retrieved cross-file context.
    \item Across all analyzed samples, the oracle agent provides \textbf{helpful context in 100\% of cases}, whereas RepoMap succeeds in surfacing helpful context in only \textbf{16.7\% of cases}.
\end{itemize}

The distribution of reasons behind each method's successful completions is shown in \Cref{fig: repomap_vs_oracle_breakdown}.

Overall, the oracle agent's advantage lies in consistently retrieving semantically aligned, high-quality cross-file context that directly drives successful function completions. RepoMap, by contrast, succeeds mainly in cases where logic errors happen to be avoided, rather than because it surfaced helpful context. These results underscore that effective context retrieval is the decisive factor behind the oracle agent's superior performance.

In the next subsection, we provide detailed case studies illustrating RepoMap's common failure modes and explaining why it often fails to surface the most relevant context.
\begin{figure}[htbp]
    \centering
    \begin{subfigure}[b]{\columnwidth}
        \centering
        \begin{adjustbox}{max width=\textwidth}
            \begin{tikzpicture}
                \pie[sum=auto, text=legend]{37/logic error, 27/context quality, 11/wrong API usage, 9/missing imports, 5/missing information}
            \end{tikzpicture}
        \end{adjustbox}
        \caption{Oracle agent: most successes are driven by high-quality retrieved context and avoiding logic errors.}
    \end{subfigure}
    \hfill
    \begin{subfigure}[b]{\columnwidth}
        \centering
        \begin{adjustbox}{max width=\textwidth}
            \begin{tikzpicture}
                \pie[sum=auto, text=legend]{31/logic error, 12/wrong API usage, 1/context quality, 3/missing information, 1/missing imports}
            \end{tikzpicture}
        \end{adjustbox}
        \caption{RepoMap: most successes stem from logic-related issues, with almost no contribution from retrieved context.}
    \end{subfigure}
    \caption{Comparison of success factors for RepoMap and the oracle agent. The oracle agent's completions are typically enabled by relevant cross-file context, while RepoMap's successes are rarely context-driven.}
    \label{fig: repomap_vs_oracle_breakdown}
\end{figure}

\subsubsection{Insights}

In this subsection, we provide a systematic analysis of why RepoMap often fails to retrieve the most useful contextual information, and how the oracle agent overcomes these limitations. Each insight is supported by concrete examples drawn from our experiments. These examples highlight the importance of retrieving semantically aligned, example-rich, and contextually relevant information for code completion. Whereas RepoMap primarily relies on API structures of imported files, which leads to systematic blind spots, the oracle agent provides richer evidence such as implementations, usage examples, conventions, and type annotations. Together, these findings underscore the benefits of employing an intelligent retrieval agent over traditional API-centric methods.
\begin{enumerate}
    \item \textbf{Insight}: RepoMap includes only function signatures without examples of usage.  
    Without usage patterns, the code completion model cannot infer how arguments should be passed or how results are handled. This gap often leads to incorrect or incomplete implementations.

    \textbf{Example}: In repository ID 30 from more-itertools, the target function is \texttt{make\_decorator}, which inserts the result of the user function at the specified \texttt{result\_index} position in the argument list passed to \texttt{wrapping\_func}:
    \begin{minted}[fontsize=\scriptsize, breaklines, frame=single, bgcolor=gray!5, escapeinside=||, breakanywhere]{python}
def make_decorator(wrapping_func, |\bfseries{|result_index|}|=0):
    def decorator_factory(*args, **kwargs):
        def decorator(user_func):
            def wrapper(*user_args, **user_kwargs):
                result = user_func(*user_args, **user_kwargs)
                args_list = list(args)
                args_list.insert(|\bfseries{|result_index|}|, result)
                return wrapping_func(*args_list, **kwargs)
            return wrapper
        return decorator
    return decorator_factory
    \end{minted}

    The RepoMap context was missing examples demonstrating how \texttt{result\_index} should be applied:
    \begin{minted}[fontsize=\scriptsize, breaklines, frame=single, bgcolor=gray!5, breakanywhere]{python}
# more_itertools/recipes.py
__all__
_marker
_sumprod
def take(n, iterable):
def tabulate(function, start=0):
def tail(n, iterable):
def consume(iterator, n=None):
def nth(iterable, n, default=None):
def all_equal(iterable, key=None):
def quantify(iterable, pred=bool):
def pad_none(iterable):
padnone
def ncycles(iterable, n):
def dotproduct(vec1, vec2):
def flatten(listOfLists):
def repeatfunc(func, times=None, *args):
def _pairwise(iterable):
def pairwise(iterable):
class UnequalIterablesError(ValueError):
    def __init__(self, details=None):
def _zip_equal_generator(iterables):
def _zip_equal(*iterables):
def grouper(iterable, n, incomplete='fill', fillvalue=None):
def roundrobin(*iterables):
def partition(pred, iterable):
def powerset(iterable):
def unique_everseen(iterable, key=None):
def unique_justseen(iterable, key=None):
def unique(iterable, key=None, reverse=False):
def iter_except(func, exception, first=None):
def first_true(iterable, default=None, pred=None):
def random_product(*args, repeat=1):
def random_permutation(iterable, r=None):
def random_combination(iterable, r):
def random_combination_with_replacement(iterable, r):
def nth_combination(iterable, r, index):
def prepend(value, iterator):
def convolve(signal, kernel):
def before_and_after(predicate, it):
    def true_iterator():
def triplewise(iterable):
def _sliding_window_islice(iterable, n):
def _sliding_window_deque(iterable, n):
def sliding_window(iterable, n):
def subslices(iterable):
def polynomial_from_roots(roots):
def iter_index(iterable, value, start=0, stop=None):
def sieve(n):
def _batched(iterable, n, *, strict=False):
def batched(iterable, n, *, strict=False):
def transpose(it):
def reshape(matrix, cols):
def matmul(m1, m2):
def factor(n):
def polynomial_eval(coefficients, x):
def sum_of_squares(it):
def polynomial_derivative(coefficients):
def totient(n):
    \end{minted}
    
    Consequently, the model's completion ignored the \texttt{result\_index} argument entirely. In contrast, the oracle agent retrieved highly relevant test cases and type annotations that illustrated the expected behavior of \texttt{result\_index}:
    \begin{minted}[fontsize=\scriptsize, breaklines, frame=single, bgcolor=gray!5, breakanywhere]{markdown}
**Explanation**: Based on the examples, the `make_decorator` function creates a multi-level wrapper:

1. First it takes a function that processes iterables
2. Returns a function that takes arguments for that function
3. Which returns a decorator that takes a user function
4. Which returns a wrapped function that applies the iterable processor to the user function's result

```
make_decorator(wrapping_func) -> decorator_factory
decorator_factory(*args, **kwargs) -> decorator
decorator(user_function) -> wrapped_function
wrapped_function(*user_args, **user_kwargs) -> processed_result
```

**Explanation**: From the tests, the `result_index` parameter determines where in the arguments to `wrapping_func` the iterable (the function's result) should be placed. By default it's the first argument (position 0):

```python
slicer = mi.make_decorator(islice)  # result_index=0
```

But it can be changed:

```python 
stringifier = mi.make_decorator(stringify, result_index=1)  # result_index=1
```

This means when stringifier is called, the result of the decorated function will be placed as the second argument to stringify.
    \end{minted}
    \item \textbf{Insight}: RepoMap fails to capture repository-specific conventions. Many codebases follow implicit conventions (e.g., returning iterators instead of lists) that guide correct implementations. RepoMap's API-only approach omits this contextual knowledge.

    \textbf{Example}: In Example~\ref{it: bm25_vs_oracle_insights_conventions} of the BM25 versus oracle comparison in \Cref{app: bm25_vs_oracle_insights}, the target function should return an iterator rather than a list. RepoMap retrieved only bare signatures, with no hints of this convention:
    \begin{minted}[fontsize=\scriptsize, breaklines, frame=single, bgcolor=gray!5, breakanywhere]{python}
# more_itertools/recipes.py
__all__
_marker
_sumprod
def take(n, iterable):
def tabulate(function, start=0):
def tail(n, iterable):
def consume(iterator, n=None):
def nth(iterable, n, default=None):
def all_equal(iterable, key=None):
def quantify(iterable, pred=bool):
def pad_none(iterable):
padnone
def ncycles(iterable, n):
def dotproduct(vec1, vec2):
def flatten(listOfLists):
def repeatfunc(func, times=None, *args):
def _pairwise(iterable):
def pairwise(iterable):
class UnequalIterablesError(ValueError):
    def __init__(self, details=None):
def _zip_equal_generator(iterables):
def _zip_equal(*iterables):
def grouper(iterable, n, incomplete='fill', fillvalue=None):
def roundrobin(*iterables):
def partition(pred, iterable):
def powerset(iterable):
def unique_everseen(iterable, key=None):
def unique_justseen(iterable, key=None):
def unique(iterable, key=None, reverse=False):
def iter_except(func, exception, first=None):
def first_true(iterable, default=None, pred=None):
def random_product(*args, repeat=1):
def random_permutation(iterable, r=None):
def random_combination(iterable, r):
def random_combination_with_replacement(iterable, r):
def nth_combination(iterable, r, index):
def prepend(value, iterator):
def convolve(signal, kernel):
def before_and_after(predicate, it):
    def true_iterator():
def triplewise(iterable):
def _sliding_window_islice(iterable, n):
def _sliding_window_deque(iterable, n):
def sliding_window(iterable, n):
def subslices(iterable):
def polynomial_from_roots(roots):
def iter_index(iterable, value, start=0, stop=None):
def sieve(n):
def _batched(iterable, n, *, strict=False):
def batched(iterable, n, *, strict=False):
def transpose(it):
def reshape(matrix, cols):
def matmul(m1, m2):
def factor(n):
def polynomial_eval(coefficients, x):
def sum_of_squares(it):
def polynomial_derivative(coefficients):
def totient(n):
    \end{minted}

    By contrast, the oracle agent retrieved contexts containing type annotations and similar functions adhering to this convention, enabling the model to generate the correct completion.
    \item \textbf{Insight}: RepoMap does not expose error-handling patterns. Without examples of defensive programming, the model tends to omit necessary checks.

    \textbf{Example}: In the previous example, the target function required handling empty input iterators. RepoMap's retrieved API structures contained no such examples, leading the model to miss critical error handling code:
    \begin{minted}[fontsize=\scriptsize, breaklines, frame=single, bgcolor=gray!5, breakanywhere]{python}
length = len(iterable)
if length == 0:
    return [iter([]) for _ in range(n)]
    \end{minted}

    In contrast, the oracle agent retrieved functions that explicitly implemented error checks, guiding the model to produce a robust completion:
    \begin{minted}[fontsize=\scriptsize, breaklines, frame=single, bgcolor=gray!5, breakanywhere]{markdown}
**Explanation**: The `split_before` function shows how to split an iterable into lists based on a predicate

From `more_itertools/more_itertools/more.py`:
```python
def split_before(iterable, pred, maxsplit=-1):
    """Yield lists of items from *iterable*, where each list ends just before
    an item for which callable *pred* returns ``True``:

        >>> list(split_before('OneTwo', lambda s: s.isupper()))
        [['O', 'n', 'e'], ['T', 'w', 'o']]

        >>> list(split_before(range(10), lambda n: n % 3 == 0))
        [[0, 1, 2], [3, 4, 5], [6, 7, 8], [9]]

    At most *maxsplit* splits are done. If *maxsplit* is not specified or -1,
    then there is no limit on the number of splits:

        >>> list(split_before(range(10), lambda n: n % 3 == 0, maxsplit=2))
        [[0, 1, 2], [3, 4, 5], [6, 7, 8, 9]]
    """
    if maxsplit == 0:
        yield list(iterable)
        return

    buf = []
    it = iter(iterable)
    for item in it:
        if pred(item) and buf:
            yield buf
            if maxsplit == 1:
                yield [item] + list(it)
                return
            buf = []
            maxsplit -= 1
        buf.append(item)
    if buf:
        yield buf
```
    \end{minted}
    \item \textbf{Insight}: RepoMap provides interfaces without implementations, docstrings, or usages.  
    As a result, the code completion model may misinterpret function purposes and misuse them.

    \textbf{Example}: In Example~\ref{it: bm25_vs_oracle_insights_semantically_similar}, RepoMap provided only the signatures of \texttt{unique\_justseen} and \texttt{unique\_everseen}, without showing how they differ. The oracle agent's contexts included implementations and usage patterns, clarifying that one checks against only the previous element while the other checks against all previously seen elements. The RepoMap-guided completion misunderstood this distinction:
    \begin{minted}[fontsize=\scriptsize, breaklines, frame=single, bgcolor=gray!5, breakanywhere]{python}
def classify_unique(iterable, key=None):
    seen = set()
    prev = object()
    use_key = key is not None
    for element in iterable:
        k = key(element) if use_key else element
        if use_key:
            try:
                if k in seen:
                    yield (element, False, False)
                else:
                    seen.add(k)
                    yield (element, True, True)
            except TypeError:
                if k in seen:
                    yield (element, False, False)
                else:
                    seen.add(k)
                    yield (element, True, True)
        else:
            if element in seen:
                yield (element, False, False)
            else:
                seen.add(element)
                yield (element, True, True)
    \end{minted}
    \item \textbf{Insight}: RepoMap does not prioritize relevance. It retrieves generic API structures without ranking them by contextual similarity, while implementations of related functions are often crucial for guiding completion.

    \textbf{Example}: In repository ID 16 from the Sympy repository, the target function was \texttt{to\_loads} in the \texttt{ForceActuator} class. RepoMap retrieved unrelated signatures, whereas the oracle agent surfaced highly relevant contexts such as the abstract method \texttt{PathwayBase.to\_loads} and examples of other actuator classes implementing their \texttt{to\_loads} methods, leading to a correct completion:
    \begin{minted}[fontsize=\scriptsize, breaklines, frame=single, bgcolor=gray!5, breakanywhere]{python}
def to_loads(self):
    return self.pathway.to_loads(self.force)
    \end{minted}
    
    These examples helped the model generate the correct completion, while RepoMap's results lacked actionable guidance:
    \begin{minted}[fontsize=\scriptsize, breaklines, frame=single, bgcolor=gray!5, breakanywhere]{python}
# sympy/abc.py
_latin
_greek
ns
_clash1
_clash2
_clash

# sympy/__init__.py
def enable_warnings():
def __sympy_debug():
SYMPY_DEBUG
test
doctest
__all__

# sympy/physics/mechanics/joint.py
__all__
class Joint(ABC):
    def __init__(self, name, parent, child, coordinates=None, speeds=None,
                 parent_point=None, child_point=None, parent_interframe=None,
                 child_interframe=None, parent_axis=None, child_axis=None,
                 parent_joint_pos=None, child_joint_pos=None):
    def __str__(self):
    def __repr__(self):
    def name(self):
    def parent(self):
    def child(self):
    def coordinates(self):
    def speeds(self):
    def kdes(self):
    def parent_axis(self):
    def child_axis(self):
    def parent_point(self):
    def child_point(self):
    def parent_interframe(self):
    def child_interframe(self):
    def _generate_coordinates(self, coordinates):
    def _generate_speeds(self, speeds):
    def _orient_frames(self):
    def _set_angular_velocity(self):
    def _set_linear_velocity(self):
    def _to_vector(matrix, frame):
    def _axis(ax, *frames):
    def _choose_rotation_axis(frame, axis):
    def _create_aligned_interframe(frame, align_axis, frame_axis=None,
                                   frame_name=None):
    def _generate_kdes(self):
    def _locate_joint_pos(self, body, joint_pos, body_frame=None):
    def _locate_joint_frame(self, body, interframe, body_frame=None):
    def _fill_coordinate_list(self, coordinates, n_coords, label='q', offset=0,
                              number_single=False):
        def create_symbol(number):
class PinJoint(Joint):
    def __init__(self, name, parent, child, coordinates=None, speeds=None,
                 parent_point=None, child_point=None, parent_interframe=None,
                 child_interframe=None, parent_axis=None, child_axis=None,
                 joint_axis=None, parent_joint_pos=None, child_joint_pos=None):
    def __str__(self):
    def joint_axis(self):
    def _generate_coordinates(self, coordinate):
    def _generate_speeds(self, speed):
    def _orient_frames(self):
    def _set_angular_velocity(self):
    def _set_linear_velocity(self):
class PrismaticJoint(Joint):
    def __init__(self, name, parent, child, coordinates=None, speeds=None,
                 parent_point=None, child_point=None, parent_interframe=None,
                 child_interframe=None, parent_axis=None, child_axis=None,
                 joint_axis=None, parent_joint_pos=None, child_joint_pos=None):
    def __str__(self):
    def joint_axis(self):
    def _generate_coordinates(self, coordinate):
    def _generate_speeds(self, speed):
    def _orient_frames(self):
    def _set_angular_velocity(self):
    def _set_linear_velocity(self):
class CylindricalJoint(Joint):
    def __init__(self, name, parent, child, rotation_coordinate=None,
                 translation_coordinate=None, rotation_speed=None,
                 translation_speed=None, parent_point=None, child_point=None,
                 parent_interframe=None, child_interframe=None,
                 joint_axis=None):
    def __str__(self):
    def joint_axis(self):
    def rotation_coordinate(self):
    def translation_coordinate(self):
    def rotation_speed(self):
    def translation_speed(self):
    def _generate_coordinates(self, coordinates):
    def _generate_speeds(self, speeds):
    def _orient_frames(self):
    def _set_angular_velocity(self):
    def _set_linear_velocity(self):
class PlanarJoint(Joint):
    def __init__(self, name, parent, child, rotation_coordinate=None,
                 planar_coordinates=None, rotation_speed=None,
                 planar_speeds=None, parent_point=None, child_point=None,
                 parent_interframe=None, child_interframe=None):
    def __str__(self):
    def rotation_coordinate(self):
    def planar_coordinates(self):
    def rotation_speed(self):
    def planar_speeds(self):
    def rotation_axis(self):
    def planar_vectors(self):
    def _generate_coordinates(self, coordinates):
    def _generate_speeds(self, speeds):
    def _orient_frames(self):
    def _set_angular_velocity(self):
    def _set_linear_velocity(self):
class SphericalJoint(Joint):
    def __init__(self, name, parent, child, coordinates=None, speeds=None,
                 parent_point=None, child_point=None, parent_interframe=None,
                 child_interframe=None, rot_type='BODY', amounts=None,
                 rot_order=123):
    def __str__(self):
    def _generate_coordinates(self, coordinates):
    def _generate_speeds(self, speeds):
    def _orient_frames(self):
    def _set_angular_velocity(self):
    def _set_linear_velocity(self):
class WeldJoint(Joint):
    def __init__(self, name, parent, child, parent_point=None, child_point=None,
                 parent_interframe=None, child_interframe=None):
    def __str__(self):
    def _generate_coordinates(self, coordinate):
    def _generate_speeds(self, speed):
    def _orient_frames(self):
    def _set_angular_velocity(self):
    def _set_linear_velocity(self):

# sympy/physics/mechanics/loads.py
__all__
class LoadBase(ABC, namedtuple('LoadBase', ['location', 'vector'])):
    def __add__(self, other):
    def __mul__(self, other):
class Force(LoadBase):
    def __new__(cls, point, force):
    def __repr__(self):
    def point(self):
    def force(self):
class Torque(LoadBase):
    def __new__(cls, frame, torque):
    def __repr__(self):
    def frame(self):
    def torque(self):
def gravity(acceleration, *bodies):
def _parse_load(load):

# sympy/physics/mechanics/pathway.py
__all__
class PathwayBase(ABC):
    def __init__(self, *attachments):
    def attachments(self):
    def attachments(self, attachments):
    def length(self):
    def extension_velocity(self):
    def to_loads(self, force):
    def __repr__(self):
class LinearPathway(PathwayBase):
    def __init__(self, *attachments):
    def length(self):
    def extension_velocity(self):
    def to_loads(self, force):
class ObstacleSetPathway(PathwayBase):
    def __init__(self, *attachments):
    def attachments(self):
    def attachments(self, attachments):
    def length(self):
    def extension_velocity(self):
    def to_loads(self, force):
class WrappingPathway(PathwayBase):
    def __init__(self, attachment_1, attachment_2, geometry):
    def geometry(self):
    def geometry(self, geometry):
    def length(self):
    def extension_velocity(self):
    def to_loads(self, force):
    def __repr__(self):
def _point_pair_relative_position(point_1, point_2):
def _point_pair_length(point_1, point_2):
def _point_pair_extension_velocity(point_1, point_2):

# sympy/physics/mechanics/rigidbody.py
__all__
class RigidBody(BodyBase):
    def __init__(self, name, masscenter=None, frame=None, mass=None,
                 inertia=None):
    def __repr__(self):
    def frame(self):
    def frame(self, F):
    def x(self):
    def y(self):
    def z(self):
    def inertia(self):
    def inertia(self, I):
    def central_inertia(self):
    def central_inertia(self, I):
    def linear_momentum(self, frame):
    def angular_momentum(self, point, frame):
    def kinetic_energy(self, frame):
    def set_potential_energy(self, scalar):
    def parallel_axis(self, point, frame=None):

# sympy/physics/vector/__init__.py
__all__
    \end{minted}
    \item \textbf{Insight}: RepoMap shows function signatures but not how they are used. This omission can lead the model to pass arguments of the right type but with incorrect values.

    \textbf{Example}: In repository ID 30 from Flask, the ground-truth function relied on the \texttt{aborter} method of the \texttt{current\_app} object:
    \begin{minted}[fontsize=\scriptsize, breaklines, frame=single, bgcolor=gray!5, escapeinside=||, breakanywhere]{python}
from .globals import current_app

def abort(code, *args, **kwargs):
    if current_app:
        current_app.|\bfseries{|aborter|}|(code, *args, **kwargs)
    _wz_abort(code, *args, **kwargs)
    \end{minted}

    RepoMap retrieved only the variable name, with no type or member function information:
    \begin{minted}[fontsize=\scriptsize, breaklines, frame=single, bgcolor=gray!5, escapeinside=||, breakanywhere]{python}
# src/flask/globals.py
_no_app_msg
_cv_app
app_ctx
|\bfseries{|current_app|}|
g
_no_req_msg
_cv_request
request_ctx
request
session
    \end{minted}
    As a result, the completion failed. The oracle agent, by contrast, retrieved definitions and usage examples of \texttt{current\_app}, including its \texttt{aborter} member, enabling the correct implementation:
    \begin{minted}[fontsize=\scriptsize, breaklines, frame=single, bgcolor=gray!5, breakanywhere]{markdown}
**Explanation**: The `redirect` function from `src/flask/helpers.py` follows a similar pattern to the `abort` function

The `redirect` function in `helpers.py` shows the pattern you need to follow for the `abort` function. It checks if `current_app` exists and delegates to it, otherwise it falls back to Werkzeug directly:

```python
# src/flask/helpers.py
def redirect(
    location: str, code: int = 302, Response: type[BaseResponse] | None = None
) -> BaseResponse:
    """Create a redirect response object.

    If :data:`~flask.current_app` is available, it will use its
    :meth:`~flask.Flask.redirect` method, otherwise it will use
    :func:`werkzeug.utils.redirect`.

    :param location: The URL to redirect to.
    :param code: The status code for the redirect.
    :param Response: The response class to use. Not used when
        ``current_app`` is active, which uses ``app.response_class``.

    .. versionadded:: 2.2
        Calls ``current_app.redirect`` if available instead of always
        using Werkzeug's default ``redirect``.
    """
    if current_app:
        return current_app.redirect(location, code=code)

    return _wz_redirect(location, code=code, Response=Response)
```

**Explanation**: The `current_app` object from `src/flask/globals.py`

The `current_app` object is imported from `globals.py` and is used to check if we are in an application context:

```python
# src/flask/globals.py
from __future__ import annotations

import typing as t
from contextvars import ContextVar

from werkzeug.local import LocalProxy

if t.TYPE_CHECKING:  # pragma: no cover
    from .app import Flask
    from .ctx import _AppCtxGlobals
    from .ctx import AppContext
    from .ctx import RequestContext
    from .sessions import SessionMixin
    from .wrappers import Request


_no_app_msg = """\
Working outside of application context.

This typically means that you attempted to use functionality that needed
the current application. To solve this, set up an application context
with app.app_context(). See the documentation for more information.\
"""
_cv_app: ContextVar[AppContext] = ContextVar("flask.app_ctx")
app_ctx: AppContext = LocalProxy(  # type: ignore[assignment]
    _cv_app, unbound_message=_no_app_msg
)
current_app: Flask = LocalProxy(  # type: ignore[assignment]
    _cv_app, "app", unbound_message=_no_app_msg
)
```
    \end{minted}
    \item \textbf{Insight}: RepoMap does not handle multi-hop dependencies. It captures only one-hop function signatures and misses deeper chains of dependency.

    \textbf{Example}: In the same Flask case, RepoMap retrieved the \texttt{current\_app} variable but completely missed its member function \texttt{aborter}, a two-hop dependency critical for correct completion.
\end{enumerate}


\section{Hyperparameter settings}\label{app: hyperparams}

This section summarizes the main hyperparameter settings used in the experiments. Unless otherwise noted, the same settings are used for both completion backbones and across the main and appendix experiments.

\paragraph{Completion models and decoding.}
We evaluate two completion backbones, Qwen3-8B and Qwen3-30B-A3B \citep{qwen3}. For generation, we use a maximum completion length of 4096 tokens, temperature 0.7, and nucleus sampling with top-$p=0.8$.

\paragraph{Agent backbones.}
In the main experiments, all agents use Claude 3.7 Sonnet \citep{claude-3-7} as the backbone. In \Cref{app: qwen3_coder}, we repeat the indexing-time agent experiments with Qwen3-Coder \citep{qwen3coder} while keeping the remaining settings unchanged.

\paragraph{Prompt context budgets.}
The left context, right context, and retrieved cross-file context are each capped at 10K tokens, as noted in \Cref{sec: exp_setup}. These budgets are applied uniformly across baselines and agent-based methods.

\paragraph{Agent budgets.}
For agent-based methods, the exploration budget is limited to 10 high-priority files per instance. The retriever, forecaster, SpecAgent, and oracle agent each return at most 12 context blocks. In the main SpecAgent configuration, these 12 blocks are composed of 9 retrieved blocks and 3 predicted blocks. The composition ablation in \Cref{sec: exp_ablation} varies the number of prediction blocks over \{0, 1, 3, 6, 12\}, with the number of retrieval blocks adjusted so that the total remains 12.

\paragraph{Sparse and dense retrieval settings.}
For retrieval baselines, repository files are segmented into 50-line chunks. The sparse retrieval query is formed from the final 50 non-empty lines of the target prompt. Reranking is performed over at most 200 candidate chunks drawn from up to 1000 candidate files, and the final prompt includes up to 5 retrieved cross-file chunks. BM25 is the default sparse ranking function. Dense retrieval uses the same candidate generation pipeline, with either UniXcoder \citep{unixcoder} or CodeSage v2 large \citep{codesage} as the reranker.

\paragraph{RepoMap combination setting.}
For the BM25 + RepoMap hybrid, the interpolation weight on the RepoMap component is set to 0.3.

\paragraph{Serving setup.}
All experiments are run on a cluster with 8 A100 GPUs, matching the setup described in \Cref{sec: exp_setup}. When batched serving is used, tensor parallelism is set to 8 so that a model is sharded across the 8 GPUs.


\section{Potential risks}

While SpecAgent introduces a practical and efficient framework for repository-aware code completion, several potential risks and considerations remain.  

\paragraph{Repository privacy and data security.}  
Indexing-time exploration involves broad repository access, which could expose sensitive information if applied to private codebases without proper access control or data governance. Deployments should ensure strict adherence to organizational privacy policies and perform indexing in secure, access-controlled environments.

\paragraph{Speculative generation reliability.}  
SpecAgent's speculative predictions, while beneficial for recall and coverage, may occasionally introduce misleading or obsolete context if the actual repository evolution diverges significantly from the predicted trajectory. Ensuring proper validation, caching strategies, and developer oversight is important to mitigate hallucinated or stale suggestions.

\paragraph{Benchmark generalization.}  
Although we design a leakage-free synthetic benchmark to avoid future context contamination, synthetic data may not perfectly represent the complexity or noise of real-world repositories. Performance reported here should therefore be interpreted as indicative of potential gains rather than definitive real-world accuracy.

\paragraph{Computational and environmental costs.}  
Indexing-time agents perform extensive repository analysis and speculative computation. While amortized over future completions, large-scale indexing may still incur nontrivial compute and energy overhead. Future work should explore more efficient incremental indexing pipelines to reduce the environmental footprint.

\paragraph{Overreliance on automation.}  
SpecAgent's proactive design could encourage overreliance on automated code suggestions. Developers should treat generated completions as assistive rather than authoritative, maintaining human oversight to ensure correctness, security, and maintainability.


\section{Licenses and responsible use}

The REPOCOD dataset \citep{repocod} is distributed under the BSD-3-Clause license.  
The Qwen3 and Qwen3-Coder models \citep{qwen3,qwen3coder} are released under the Apache License 2.0.  
All other third-party tools and libraries used in this work comply with their respective open-source licenses.  
No proprietary or restricted data sources were used in our experiments.

\paragraph{Consistency with intended use.}  
All artifacts used in this study were employed in a manner consistent with their stated intended use.  
REPOCOD was explicitly released for research on repository-level code understanding and generation, aligning with our use in evaluating retrieval and completion methods.  
Similarly, Qwen3 and Qwen3-Coder are open-source LLMs intended for research and development purposes, and our use was confined strictly to academic experimentation and analysis.  
All derived datasets and benchmarks introduced in this work are designed solely for research and evaluation under the same conditions, and are not intended for production or commercial deployment.

\paragraph{Data privacy and content verification.}  
The REPOCOD dataset is derived from publicly available open-source repositories and does not contain personally identifying information or private code under restrictive licenses.  
Before conducting experiments, we verified that no data in our evaluation corpus contained names, credentials, or other sensitive identifiers.  
Our synthetic benchmark, created by applying the function removal process to REPOCOD, inherits this property and introduces no additional human-related or offensive content.  
No human annotation, crowdsourced labeling, or user-generated personal data was collected or processed in this study.

\paragraph{Anonymization and protection.}  
All intermediate artifacts (e.g., indexed repository states and context blocks) were stored and analyzed on secure research servers with controlled access.  
No attempt was made to de-anonymize contributors to the original repositories, and no identifiers linking to individual developers were used in training or evaluation.

\end{document}